\documentstyle[aps,pre]{revtex}

\begin{document}

\renewcommand{\baselinestretch}{1.25}

\draft

\title{Notes on Brownian motion and related phenomena}

\author{Deb Shankar Ray\footnote{e-mail : pcdsr@mahendra.iacs.res.in} }

\address{Department of Physical Chemistry, Indian Association for the
Cultivation of Science, Jadavpur, Calcutta 700032, India.}

\date{\today}

\maketitle

\begin{abstract}
In this article we explore the phenomena of nonequilibrium stochastic process 
starting from the phenomenological Brownian motion. The essential points are 
described in terms of Einstein's theory of Brownian motion and then the theory 
extended to Langevin and Fokker-Planck formalism. Then the theory is applied 
to barrier crossing dynamics, popularly known as Kramers' theory of activated 
rate processes. The various regimes are discussed extensively and Smoluchowski 
equation is derived as a special case. Then we discuss some of the aspects of 
Master equation and two of its applications. 
\end{abstract}


\newpage

\tableofcontents

\newpage

\section{Brownian motion : Einstein's theory}

\subsection{Introduction}

The so-called Brownian motion was described for the first time in the year 
1828 by the botanist Robert Brown. In investigating the pollen of different 
plants he observed that this become dispersed in water in a great number of 
small particles - the pollen grains. These were perceived to be in 
uninterrupted and irregular swarming motion. As the phenomenon repeated 
itself with all kinds of organic substances, he thought that he had found in 
these particles the `primitive molecule' of living matter. He also found that 
all kinds of inorganic substances presented the same phenomenon and drew the 
conclusion that all matter was built up of `primitive molecules'.

Of the authors who carried out investigations on the Brownian movement before 
Einstein, we mention : Regnault (1858), Weiner (1863), Jevons (1870), Dancer 
(1870) and Delsaux (1877).

The first precise investigation was due to Gouy (1888), who observed that the 
motion is more lively the smaller the viscosity of the liquid. He also 
ascribed the motion to the effect of thermal molecular motion of the liquid. 
Besides Gouy's work there was only other investigation of precise nature by 
Exner (1900) who showed that the velocity of the movement decreases with the 
size of the particle and increases with rise of temperature.

Einstein was the first (1905) to formulate a correct picture of the entire 
problem. We discuss his theory in Sec. 1.2, 1.3, 1.4. and 1.6. {\footnote{
We draw heavily from the two classic papers I and IV of Ref.\cite{furth} } }

\subsection{The irregular movement of particles suspended in a liquid and its 
relation to diffusion}

(a) Suppose there be suspended particles irregularly dispersed in a liquid. 
The particles are of microscopically visible size and their movements are of
such magnitude that they can be observed under a microscope. We first consider
here the irregular movements of particles which arise from {thermal molecular}
movement. This gives rise to diffusion.

Evidently it must be assumed (i) that each particle executes a movement which
is independent of movement of all other particles. (ii) The movement of one
and same particle after different intervals of time must be considered as
mutually independent processes, if we think these intervals are not too small.

We introduce a time interval $\tau$, which is very small compared to observed
interval of time, but at the same time large such that the motion executed by 
a particle in two consecutive intervals $\tau$ are mutually independent.

Suppose that there are $n$ suspended particles. In an interval $\tau$, 
$x$-coordinate of the single particle will increase by $\Delta$. $\Delta$ has
a different (positive or negative) value for each particle. For the value of
$\Delta$ a certain probability law will hold. Let $dn$ be the number of
particles which experience a displacement between $\Delta$ and $\Delta+d\Delta$
in time interval $\tau$. Then,
\begin{equation}
\frac{dn}{n} = \phi(\Delta) \; d\Delta
\end{equation}

\noindent
where the total probability is one, i.e.,
\begin{equation}
\int_{-\infty}^{+\infty} \phi(\Delta) \; d\Delta = 1 \; \; .
\end{equation}

\noindent
Here $\phi(\Delta)$ is the probability of jump of magnitude $\Delta$ for the
particle, $\phi$ only differs from zero for very small values of $\Delta$ and
fulfills the condition,
\begin{equation}
\phi(\Delta) = \phi(-\Delta) \; \; .
\end{equation}

We confine ourselves to motion in one dimension ($x$). Let $\nu = f(x,t)$, 
the number of particles per unit volume. We now calculate the distribution of
particles at time $t+\tau$ from a distribution at time $t$.

\noindent
We consider two planes perpendicular to $x$ axis at $x$ and $x+\Delta$. Then
the distribution at time $t$ and space $x+\Delta$ evolves to a distribution
at time $t+\tau$ and at $x$ as follows,
\begin{equation}
f(x,t+\tau) = \int_{-\infty}^{+\infty} f (x+\Delta,t) \; \phi(\Delta) \;
d\Delta
\end{equation}

\noindent
The integration over $\Delta$ takes into account of all possible jumps from 
$x+\Delta$ to $x$ each with a probability $\phi(\Delta)$.

\noindent
Since $\tau$ is small we write
\begin{equation}
f(x,t+\tau) = f(x,t) + \tau \; \frac{\partial f}{\partial t}
\end{equation}

\noindent
Again since $\Delta$ is small
\begin{equation}
f(x+\Delta,t) = f(x,t) + \Delta \; \frac{\partial f(x,t)}{\partial x} +
\frac{\Delta^2}{2!} \; \frac{\partial^2 f(x,t)}{\partial x^2} + \ldots
\end{equation}

\noindent
Putting (5) and (6) in (4) we obtain
\begin{eqnarray}
f+ \frac{\partial f}{\partial t} \; \tau & = & \int^{+\infty}_{-\infty} f(x,t)
\phi(\Delta) \; d\Delta + \int^{+\infty}_{-\infty} \left (
\frac{\partial f}{\partial x} \right ) \Delta \; \phi(\Delta) \; d\Delta +
\int^{+\infty}_{-\infty} \left ( \frac{\partial^2 f}{\partial x^2} \right )
\frac{\Delta^2}{2!} \; \phi(\Delta) \; d\Delta \nonumber \\
& = & f(x,t) \int^{+\infty}_{-\infty} \phi(\Delta) \; d\Delta 
+ \left ( \frac{\partial f}{\partial x} \right )  \int^{+\infty}_{-\infty} 
\Delta \; \phi(\Delta) \; d\Delta 
+ \left ( \frac{\partial^2 f}{\partial x^2} \right ) \int^{+\infty}_{-\infty} 
\frac{\Delta^2}{2!} \; \phi(\Delta) \; d\Delta
\end{eqnarray}

\noindent
Since
\begin{eqnarray*}
\int^{+\infty}_{-\infty} \phi(\Delta) \; d\Delta = 1
\end{eqnarray*}

\noindent
and {\footnote{The jump of magnitude $\Delta$ has no preferential direction}}
\begin{eqnarray*}
\phi(\Delta) = (-\Delta) \; \; , \; \; \; \; \phi(\Delta) \; 
{\rm an}\; {\rm even} \; {\rm function} \; \; , 
\end{eqnarray*}

\noindent
i.e., $\int^{+\infty}_{-\infty} \Delta \; \phi(\Delta) = 0$. We obtain
\begin{eqnarray}
\frac{\partial f}{\partial t} \; \tau & = & \frac{\partial^2 f}{\partial x^2}
\int^{+\infty}_{-\infty} \frac{\Delta^2}{2!} \; \phi(\Delta) \; d\Delta 
\nonumber \\
\frac{\partial f}{\partial t} & = & \left [ \; \frac{1}{\tau} 
\int^{+\infty}_{-\infty} \frac{\Delta^2}{2!} \; \phi(\Delta) \; d\Delta \;
\right ]  \frac{\partial^2 f}{\partial x^2}
\end{eqnarray}

\noindent
Putting
\begin{eqnarray*}
D = \frac{1}{\tau} \;
\int^{+\infty}_{-\infty} \frac{\Delta^2}{2!} \; \phi(\Delta) \; d\Delta \; \; .
\end{eqnarray*}

\noindent
We obtain the diffusion equation :
\begin{equation}
\frac{\partial f(x,t)}{\partial t} = D \; 
\frac{\partial^2 f(x,t)}{\partial x^2}
\end{equation}
\noindent
where $D$ is the diffusion coefficient.

Note that both the equations (4) and (9) express the same law of evolution of 
distribution of particles in space and time. While (4) is of integral form, 
(9) is a differential equation.

(b) Next problem that we investigate is how the distribution spreads in time.
Mathematically speaking, this is an initial value problem.

Let us suppose that at $t=0$ all the particles are concentrated at a point
$x=0$. This means that the density or the distribution function $f(x,t)$ is 
infinite at this point and zero everywhere ( except this point ). We consider 
the problem of diffusion outward from this point.
\begin{equation}
{\rm At} \; \; \; t=0 \; \; , \; \; \; \; f(x,0) = n \; \delta(x)
\end{equation}

\noindent
where by definition
\begin{eqnarray*}
\delta(x) & = & 0 \; \; , \; \; \; \; x\neq 0 \nonumber \\
\delta(x) & = & \infty \; \; , \; \; \; \; x = 0 
\end{eqnarray*}

\noindent
Also we have,
\begin{eqnarray*}
f(x_0) = \int^{+\infty}_{-\infty} f(x) \; \delta (x-x_0) \; dx \; \; .
\end{eqnarray*} 

\noindent
Fourier representation of $\delta (x)$ gives
\begin{eqnarray*}
\delta (x) = \frac{1}{2\pi} \; \int^{+\infty}_{-\infty} e^{ikx} \; dk \; \; .
\end{eqnarray*}

\noindent
Writing $f(x,t)$ as a Fourier transform,
\begin{equation}
f(x,t) = \frac{1}{2\pi} \; \int^{+\infty}_{-\infty} e^{ikx} \; 
{\tilde f}(k,t) \; dk
\end{equation}

\noindent
and putting it in Eq.(9) we have
\begin{eqnarray*}
\frac{1}{2\pi} \; \int^{+\infty}_{-\infty} e^{ikx} \;
\frac{ \partial {\tilde f}(k,t) }{\partial t} \; dk =
\frac{D}{2\pi} \; \int^{+\infty}_{-\infty} (-k^2) \; e^{ikx} \;
{\tilde f}(k,t) \; dk \; \; .
\end{eqnarray*}

\noindent
Therefore,
\begin{equation}
\frac{\partial {\tilde f}(k,t)}{\partial t} = -k^2 \; D \; {\tilde f}(k,t)
\end{equation}

\noindent
which can be solved to give
\begin{equation}
{\tilde f}(k,t) = {\tilde f}(k,0) \; e^{-k^2 \; D \; t} \; \; .
\end{equation}

\noindent
Since $f(x,0) = n\; \delta(x)$ and using the definition of $\delta(x)$ we 
have
\begin{eqnarray}
n \; \delta(x) & = & \frac{n}{2\pi} \int^{+\infty}_{-\infty} \; e^{ikx} \; dk
\nonumber \\
& = & \frac{1}{2\pi} \; \int^{+\infty}_{-\infty} e^{ikx} \; n \; dk \; \; .
\end{eqnarray}

\noindent
Again from (11)
\begin{eqnarray*}
f(x,0) = \frac{1}{2\pi} \; \int^{+\infty}_{-\infty} e^{ikx} \; 
{\tilde f}(k,0) \; dk \; \; ,
\end{eqnarray*}

\noindent
hence by (10) we have
\begin{equation}
n \; \delta (x) = \frac{1}{2\pi} \; \int^{+\infty}_{-\infty} e^{ikx} \;
{\tilde f}(k,0) \; dk \; \; .
\end{equation}

\noindent
Comparison with (14) gives
\begin{eqnarray*}
{\tilde f} (k,0) = n \; \; .
\end{eqnarray*}

\noindent
Therefore,
\begin{equation}
{\tilde f}(k,t) = n \; e^{-k^2 \; D \; t} \; \; .
\end{equation}

\noindent
Putting (16) in (11) we get the distribution at t [ see Appendix-A for details ]

\begin{equation}
f(x,t) = \frac{n}{ \sqrt{4\pi \; D \; t} } \; e^{-x^2/4 \; D \; t}
\end{equation}

\noindent
The above equation shows how the distribution spreads in time (as shown
in the figures ),

(c) We now calculate the displacement $\lambda_x$ in the direction of $x$-axis 
which a single particle experiences on an average - more accurately expressed -
the square root of the arithmetic mean of the squares of the displacement,
\begin{eqnarray}
\langle x^2 \rangle  & = & \frac{1}{n} \; \int^{+\infty}_{-\infty}
x^2 \; f(x,t) \; dx \nonumber \\
& = & \int^{+\infty}_{-\infty} x^2 \; 
\frac{1}{ \sqrt{4\pi \; D \; t} } \; e^{-x^2/4 \; D \; t} dx \nonumber \\
& = & \sqrt{ \frac{\alpha}{\pi} } \; \int^{+\infty}_{-\infty} x^2 \;
e^{-\alpha x^2} \; dx \; \; \; , \; \; \; \alpha = 1/4Dt \nonumber \\
& = & \sqrt{ \frac{\alpha}{\pi} } \; (-1) \;  \int^{+\infty}_{-\infty}
\frac{\partial}{\partial \alpha} e^{-\alpha x^2} \; dx \nonumber \\
& = & - \sqrt{ \frac{\alpha}{\pi} } \; \frac{\partial}{\partial \alpha}
\int^{+\infty}_{-\infty} e^{-\alpha x^2} \; dx \nonumber \\
& = & - \sqrt{ \frac{\alpha}{\pi} }  \frac{\partial }{\partial \alpha} \;
\sqrt{ \frac{\pi}{\alpha} } \nonumber \\
& = & \frac{1}{2 \; \alpha} \; \; .
\end{eqnarray}

\noindent
Thus $ \langle x^2 \rangle = 2Dt$ or
\begin{equation}
\lambda_x = \sqrt{ \langle x^2 \rangle } = \sqrt{2 \; D \; t} \; \; .
\end{equation}

The mean displacement is therefore proportional to the square root of time.
This is a typical characteristic of Brownian motion.

\subsection{Diffusion and mobility}

We consider the suspended particles irregularly dispersed in a liquid. We 
consider this state of dynamic equilibrium, on the assumption that a force 
$k$ acts on the particles and it depends only on position but not on time. 
For simplicity, we assume that the force is exerted everywhere in the direction 
of $x$-axis.

We can look upon the dynamic equilibrium condition as a superposition of two
processes acting in the opposite directions.

(i) A movement of the suspended particle under the influence of the external 
force $k$. To be specific we consider a particle moving under the force of
gravity $k$ in a liquid. The particle is of spherical form with a radius
$\rho$ and the liquid has a coefficient of viscosity $\kappa$. When the 
particle immersed in the liquid is falling under gravitational force $k$ it
experiences an opposing force of hydrodynamic origin. When the external force
$k$ balances the opposing force, the particle falls with a constant terminal
velocity $v_0$. Thus if the external force of gravity is $k$ and the opposing
force is $6\pi \kappa \rho v_0$, then
\begin{equation}
v_0 = \frac{k}{6 \pi \kappa \rho} \; \; .
\end{equation}

\noindent
If $\nu$ is the no. of particles per unit volume, then $\nu \; v_0$ number
of particles pass a unit area per unit time ( or $\nu k/6 \pi \kappa \rho$ )
under the action of external force $k$.

(ii) The process of diffusion which is looked upon as a result of irregular
movement of the particles produced by the thermal molecular movement of the
liquid.

If $D$ is the coefficient of diffusion of suspended and $\mu$ the mass of 
a particle, then due to diffusion, the number of particles passing per unit 
area per unit time is
\begin{eqnarray*}
-D\frac{\partial \nu}{\partial x}
\end{eqnarray*}

\noindent
( $-D\frac{\partial \mu \nu}{\partial x}$ grams of particles crossing per 
unit area per unit time )

Under the condition of dynamic equilibrium we must have
\begin{mathletters}
\begin{equation}
\frac{\nu k}{6 \pi \kappa \rho} = -D \; \frac{\partial \nu}{\partial x} \; \; .
\end{equation}

At equilibrium, since the density under the force of gravity $k$ varies as
( Boltzmann distribution )
\begin{equation}
\nu = \nu_0 \; \exp \left [ -\frac{ k(x-x_0) N}{RT} \right ]
\end{equation}
\end{mathletters}

\noindent
where $N$ is the Avogadro number and $R$ is the universal gas constant. 
Putting (21b) in (21a) we obtain,
\begin{equation}
D = \frac{RT}{N} \; \frac{1}{6 \pi \kappa \rho}
\end{equation}

\noindent
The coefficient of diffusion of the suspended particles therefore depends only 
on the coefficient of viscosity of the liquid and on the size of the suspended
particles.

If $\frac{1}{6 \pi \kappa \rho}$ is denoted $B$ ( ``mobility'' of the particle )
then (22) may be rewritten as,
\begin{equation}
D = \frac{RT}{N} \; B
\end{equation}

This is the relation between mobility of the suspended particle and the
diffusion coefficient.

\subsection{Determination of Avogadro number}

We have found the diffusion coefficient $D$ of a material suspended in a liquid
in the form of small spheres of radius $\rho$ as,
\begin{equation}
D = \frac{RT}{N} \left ( \frac{1}{6 \pi \kappa \rho} \right )
\end{equation}

Again the mean value of the displacement of the particle in $x$-direction
in time $t$,
\begin{equation}
\lambda_x = \sqrt{ 2Dt}
\end{equation}

\noindent
By eliminating $D$ we obtain
\begin{equation}
\lambda_x = \sqrt{t} \sqrt{ \frac{RT}{N} \frac{1}{3\pi \kappa \rho} }
\end{equation}

\noindent
This equation shows how $\lambda_x$ depends on $T$, $\kappa$ and $\rho$. We
will now calculate how great $\lambda_x$ is for one second. Take
\begin{eqnarray*}
N & = &  6 \times 10^{23} \\
\kappa & = & 1.35 \times 10^{-2} \; {\rm poise} \; ( {\rm water} \; {\rm at} \; 17^{\circ} c )
\; \; \; ; \; \; \; 1 \; {\rm poise} = 1 {\rm gm.cm^{-1}.sec^{-1} }  \\
\rho & = & 0.001 \; {\rm mm} \\
{\rm R} & = & 8.31 \times 10^7 \; {\rm erg.mole^{-1}.deg.K^{-1} }
\end{eqnarray*}

\noindent
We get
\begin{eqnarray*}
\lambda_x = 8 \times 10^{-5} \; {\rm cm} \; \; .
\end{eqnarray*}

\noindent
On the other hand the relation (26) can be used for determination of $N$. We
obtain thus
\begin{equation}
N = \frac{1}{ \lambda_x^2} \frac{RT}{3\pi \kappa \rho}
\end{equation}

\noindent
where $\lambda_x$ has to be determined experimentally.

\subsection{Experimental confirmation}

The first of the investigations confirming Einstein's formula in its original 
meaning was carried out by Seddig (1908) who look two photograph of an
aqueous suspension of cinnabar on the same plate at an interval of $0.14$ sec 
and measured the distance of the corresponding images on the plate. He found
that on an average the displacement at different temperatures were inversely
proportional  to the viscosities as the theory demanded.

Henri (1908) found similarly with the aid of cinematograph records of mean
displacement of particles Caoutchone that the time law, $x^2$ proportional
to $t$, was followed.

The establishment of first complete and absolute proof of the formula lies to 
the credit of Perrin and his group (1914), who followed the movements of 
single particle of gamboge or mastic under a microscope and recorded their 
positions at equal time intervals by means of an indicating apparatus. They
determined $N$ and found the value between $56$ and $88 \times 10^{22}$.

\subsection{Theoretical observations on Brownian motion and the existence of
a random force}

It is now well established that irregular movements of the suspended
particles in a liquid are caused by thermal motion of the molecules of the
liquid. We now put forward two observations on the Brownian motion from a
theoretical point of view to establish the existence of a random force.

(i) From the molecular theory of heat we can calculate the mean value of the
instantaneous velocity which the particle may have at the absolute
temperature $T$. Thus the kinetic energy of the motion of a particle is
independent of the size and nature of the particle and independent of
the nature of its environment, e.g., of the liquid in which the particle
is suspended. The mean velocity $\sqrt{ \langle v^2 \rangle }$ of the particle
of mass $m$ is therefore determined by the equation
\begin{equation}
m \frac{ \langle v^2 \rangle}{2} = \frac{3}{2} \frac{RT}{N}
\end{equation}

\noindent
with $R=8.3 \times 10^7 \; {\rm erg.mole^{-1}.deg.K^{-1} }$ and
$N=6 \times 10^{23}$.

We calculate $\sqrt{ \langle v^2 \rangle }$ for particles in colloidal
platinum solutions. for these particles we have the mass
$m=2.5 \times 10^{-15}$ gm so that for $T=292 \; {\rm K}$
\begin{equation}
\sqrt{ \langle v^2 \rangle } = \sqrt{ \frac{3RT}{mN} } = 8.6 \; {\rm cm/sec}
\end{equation}

(ii) We will now examine whether there is any prospect of actually observing
this enormous velocity of a suspended particle.

If we know nothing of the kinetic theory, we should expect the following
thing to happen.

Suppose that we impart to a particle suspended in a liquid certain velocity
$v$ by a force applied to it from outside. Then this velocity will die away
rapidly on account of the friction of the liquid. The opposing force
experienced by the particle is $6\pi \kappa \rho v$, where
$\kappa=$viscosity of the liquid, $\rho=$radius of the particle and $v$ is
the velocity of the particle. We obtain
\begin{equation}
m \frac{dv}{dt} = -6\pi \kappa \rho v \; \; ,
\end{equation}

\noindent
On integration (30) gives
\begin{equation}
v = v_0 \exp \left ( -\frac{6 \pi \kappa \rho}{m} t \right ) \; \; .
\end{equation}

\noindent
From this we calculate the time in which the velocity die away to one tenth
of its original value, i.e., $t_{1/10}$. From (31)
\begin{eqnarray}
\ln \frac{v}{v_0} & = & - \frac{6\pi \kappa \rho}{m} t_{1/10} \nonumber\\
t_{1/10} & = & \frac{\ln 10}{6\pi \kappa \rho/m} \; \; , \; \; {\rm with}
\; \; \frac{v}{v_0} = \frac{1}{10}
\end{eqnarray}

\noindent
For platinum particles ( in water ) we have put
\begin{eqnarray*}
\rho & = & 2.5 \times 10^{-6} \; {\rm cm} \\
\kappa & = & 0.01 \; {\rm poise} \\
m & = & 2.5 \times 10^{-15} \; {\rm gm}
\end{eqnarray*}

\noindent
so that we get
\begin{eqnarray*}
t_{1/10} = 3.3 \times 10^{-7} \; {\rm sec}
\end{eqnarray*}

This means that the particle nearly completely looses its original velocity
in the very short time $t_{1/10}$ through friction. But at the same time we
must assume that the particle get new impulses from the liquid molecules
during this time by some process that is the inverse of viscosity so that it
retains a velocity $\sqrt{ \langle v^2 \rangle }$ on an average. But since we
must assume that the direction and magnitude of these impulses are independent
of the original velocity and direction of motion of the particles, we must
conclude that the velocity and direction of the motion of the particle will
be greatly altered in a very short time $t_{1/10}$ and in a totally irregular
manner.

It is therefore impossible at least for ultramicroscopic particle to
ascertain $\sqrt{ \langle v^2 \rangle }$ by observation.

(iii) Although $\sqrt{ \langle v^2 \rangle }$ can not be observed, the change 
in position in time $\tau$ ( which is much larger than $t_{1/10}$ ) can be 
determined. We have already
\begin{equation}
\lambda_x = \sqrt{\tau} \; \sqrt{ \frac{RT}{N} \left ( 
\frac{1}{3\pi \kappa \rho} \right ) }
\end{equation}

\noindent
$\lambda_x$ is the change in $x$-coordinate in time $\tau$. Then the mean 
velocity in time interval $\tau$, we define as
\begin{equation}
\frac{\lambda_x}{\tau} = \frac{1}{\sqrt{\tau}} \; \sqrt{ \frac{RT}{N} \left ( 
\frac{1}{3\pi \kappa \rho} \right ) }
\end{equation}

\noindent
Since an observer can never perceive the actual path traversed in a arbitrary 
small time a certain mean velocity like $\lambda_x/\tau$ will appear to him as 
the  instantaneous velocity. This is a measurable quantity.

We have already seen that $\sqrt{ \langle v^2 \rangle }$ is enormous for a
particle when it is suspended in a liquid. On the other hand hydrodynamics 
tells us that this velocity if imparted to the particle by an impulsive force, 
will die away very rapidly. Therefore, to maintain the average 
$\sqrt{ \langle v^2 \rangle }$ the particle must experience impulses from the 
liquid molecules whose direction and magnitude are random. So to reconcile the 
kinetic theory with hydrodynamics we conclude that there must exist a random 
force $F(t)$ acting on each particle, so that we write
\begin{equation}
m \frac{dv}{dt} = - 6 \pi \kappa \rho v + F(t)
\end{equation}

\noindent
Eq.(35) is an instantaneous description of motion of the particle. Here the 
average value of the random force is zero
\begin{eqnarray*}
\langle F(t) \rangle = 0 \; \; \; \; and \; \; \; \; 
\langle F(t)^2 \rangle \neq 0
\end{eqnarray*}

\noindent
Eq.(35) is the Langevin equation of motion for the particle. We discuss this 
equation in the next section.

\section{Langevin description of Brownian motion}

\setcounter{equation}{0}

\subsection{Introduction}

In the studies on Brownian motion we are principally concerned with the 
perpetual irregular motions exhibited by small grains or particles of
colloidal size suspended in a liquid. As is now wellknown, we witness in 
Brownian movement the phenomenon of molecular agitation on a reduced scale by 
particles very large on a molecular scale - so large, in fact, as to be 
readily visible in an ultra-microscope. The perpetual motion of the Brownian 
particles is maintained by the fluctuations in the collisions with the 
molecules of the surrounding liquid. Under normal conditions in a liquid a
particle will suffer as many as $10^{21}$ collisions per second and this is
so frequent that we cannot talk about separate collisions. And it is impossible
to follow the path in any detail

In the absence of any external force, one writes the Langevin equation for
a free particle as
\begin{equation}
m\frac{dv}{dt} = -6\pi\eta r v + F(t)
\end{equation}

\noindent
$\eta$ is the viscosity of the liquid, $r$ is the radius of the particle and 
$v$ the velocity of the particle. According to this equation the influence of 
the liquid medium on the motion of the particle can be split up into two parts :

\noindent
$\bullet$ A systematic part $-6\pi\eta r v$, represents a dynamical friction 
experienced by the Brownian particle. 

\noindent
$\bullet$ A fluctuating force, $F(t)$, which is characteristic of the 
Brownian motion.

\noindent
Two assumptions are made

\begin{center}
(i) $F(t)$ is independent of $v$ and $\langle F(t) \rangle = 0$.
\end{center}

\begin{center}
(ii) $F(t)$ varies extremely rapidly compared to variations of $v$.
\end{center}

The second assumption implies that the time intervals of duration $\Delta t$
exist such that during $\Delta t$ the variations in $v$ to be very small, 
while during the same interval $F(t)$ may undergo many fluctuations. Although
$\langle F(t) \rangle$ is zero, the average $\langle F(t)^2 \rangle$ does not 
vanish since negative swings of $F(t)$ yield positive squared values. Suppose
that the minimum time in which $F(t)$ changes appreciably is called the
correlation time, $\tau_c$. The average of the product $ F(t) F(t')$,
vanishes for $|t-t'| > \tau_c$. Hence, $\langle F(t) F(t') \rangle$, the
correlation function of the random force is peaked about $t=t'$ and falls off 
to zero in a time difference $|t-t'| = \tau_c$. If $\tau_c$ is less than all
other times of interest, e.g., $1$/(damping constant), we write
\begin{equation}
\langle F(t) F(t') \rangle = 2{\cal D} \delta(t-t')
\end{equation}

\noindent
where ${\cal D}$ is some constant expressing the magnitude of the fluctuating
forces. Eq.(2) along with $\langle F(t) \rangle = 0$, completely defines the 
Langevin equation (1).

\subsection{General expression for mean square displacement}

The Langevin equation for the particle
\begin{eqnarray*}
m\frac{dv}{dt} = -6 \pi\eta r v + F(t)
\end{eqnarray*}
gives $\frac{d \langle v \rangle}{dt} = -6 \pi \eta r \langle v \rangle$
since $\langle F(t) \rangle = 0$.

Therefore,
\begin{equation}
\langle v(t) \rangle = \langle v(0) \rangle
\exp \left( - \frac{6\pi \eta r}{m} t \right)
\end{equation}

Denoting $6\pi\eta r/m = \Gamma$ (damping constant) we rewrite Eq.(1) as
\begin{equation}
m\ddot{x} = -m \Gamma \dot{x} + F(t) \; \; .
\end{equation}
since $\dot{x}=v$, therefore,
\begin{equation}
\ddot{x} = -\Gamma \dot{x} + \frac{F(t)}{m}
\end{equation}
Multiplying both sides by x we get,
\begin{equation}
x \ddot{x} = - \Gamma x \dot{x} + x \frac{F(t)}{m}
\end{equation}

\begin{eqnarray*}
{\rm Since} \; \; \; & \dot{x^2} = & 2 x \dot{x} \\
{\rm and}   \; \; \; & \ddot{x^2} = & 2(\dot{x})^2 + 2 x \ddot{x}\\
{\rm Thus} \; \; \; & x \ddot{x} = &\frac{1}{2} \ddot{x^2} - (\dot{x})^2 \; .
\end{eqnarray*}
With these relations we rewrite Eq.(6) as,
\begin{equation}
\frac{1}{2}\ddot{x^2} - (\dot{x})^2  = -\frac{\Gamma}{2}\dot{x}^2 +
\frac{1}{m} x F(t) 
\end{equation}
\noindent
or
\begin{equation}
\ddot{x^2} - 2(\dot{x})^2  =  - \Gamma \dot{x}^2 + \frac{2}{m} x F(t)
\end{equation}

Taking the average we obtain,
\begin{equation}
\frac{d^2}{dt^2} \langle x^2 \rangle - 2 \langle ( \dot{x} )^2 \rangle
= - \Gamma \frac{d}{dt} \langle x^2 \rangle + \frac{2}{m} \langle x F(t) 
\rangle
\end{equation}

\noindent
Since{\footnote{
By equipartition of energy : $\frac{1}{2} m \langle v^2 \rangle =
\frac{1}{2} k_BT$, in one dimension }}
\begin{eqnarray*}
\dot{x} = v \; \; , \; \; 2\langle (\dot{x} )^2 \rangle = 2\langle v^2 \rangle
= 2\frac{k_BT}{m}
\end{eqnarray*}

\noindent
Also{\footnote{The random force is independent of the position of the 
particle}}
\begin{eqnarray*}
\langle x F(t) \rangle = 0 \; \; .
\end{eqnarray*}

\noindent
Thus from (9) we have
\begin{equation}
\frac{d^2}{dt^2} \langle x^2 \rangle + \Gamma \frac{d}{dt} \langle x^2 \rangle 
- 2\frac{k_BT}{m} = 0 \; \; .
\end{equation}

\noindent
We now solve Eq.(10) to obtain mean square displacement $\langle x^2 \rangle$.
Put $\frac{d}{dt} \langle x^2 \rangle = y $ and $c = 2\frac{k_BT}{m}$,
then Eq.(10) gives,
\begin{equation}
\dot{y} + \Gamma y - c = 0
\end{equation}

\noindent
Let $ \Gamma y - c = y'$, then
\begin{eqnarray}
\dot{y'} + \Gamma y & = & 0 \nonumber \\
{\rm therefore} \; \; \; \; y' & = & A \exp(-\Gamma t) \; \; , \; \; 
A = {\rm constant} \; \; .
\end{eqnarray}

\noindent
At $t=0$, $y' = A$, hence
\begin{eqnarray*}
\Gamma y - c & = & A \\
{\rm or} \; \; \; \; A & = & -c \; \; ; \; \; c=2(k_BT/m)
\end{eqnarray*}

\noindent
From (12)
\begin{eqnarray*}
\Gamma y - c & = & A \exp(-\Gamma t) \\
y & = & \frac{c}{\Gamma} + \frac{A}{\Gamma } \exp(-\Gamma t)
\end{eqnarray*}

\noindent
Since, $\frac{d}{dt} \langle x^2 \rangle = y$,
\begin{eqnarray*}
\frac{d}{dt} \langle x^2 \rangle  =  \frac{c}{\Gamma} + \frac{A}{\Gamma } 
\exp(-\Gamma t)
\end{eqnarray*}

\noindent
On integration over $0$ to $t$ we obtain,
\begin{eqnarray*}
\langle x^2 \rangle  =  \frac{c}{\Gamma} t + 
\frac{A}{\Gamma^2 }  ( 1- e^{-\Gamma t} )
\end{eqnarray*}

\noindent
Since $A=-c$ and $c=2\frac{k_BT}{m}$, we obtain
\begin{equation}
\langle x^2 \rangle  =  \frac{2k_BT}{m \Gamma} t + 
\frac{2 k_BT}{m \Gamma^2 }  ( 1- e^{-\Gamma t} )
\end{equation}

\noindent
This is the general expression for mean square displacement of a Brownian 
particle suspended in a fluid.

\subsubsection{long time limit}

When $t\rightarrow \infty$, $e^{-\Gamma t}=0$, we obtain the asymptotic 
behavior,
\begin{eqnarray*}
\langle x^2 \rangle = \frac{2 k_BT}{m \Gamma} t \; \; ( {\rm constant} \;
{\rm part} \; {\rm is} \; {\rm neglected} )
\end{eqnarray*}

\noindent
Putting $\Gamma = \frac{6 \pi \eta r}{m}$, we have,
\begin{equation}
\langle x^2 \rangle = \frac{2 k_BT}{ 6 \pi \eta r} t  \; \; .
\end{equation}

\noindent
The mean square displacement is proportional to time, $t$ by (14). Comparison 
with Einstein's expression $\langle x^2 \rangle = 2Dt$ gives
\begin{eqnarray*}
D = \frac{k_BT}{6 \pi \eta r} \; \; ,
\end{eqnarray*}

\noindent
which is the standard expression for diffusion coefficient calculated 
earlier by Einstein's method.

\subsubsection{short time limit}

When $t \rightarrow$ small, from (13)
\begin{eqnarray*}
\langle x^2 \rangle  =  \frac{2k_BT}{m \Gamma} t + 
\frac{2 k_BT}{m \Gamma^2 }  ( 1- e^{-\Gamma t} )
\end{eqnarray*}

\noindent
we expand the exponential term to recover the leading order time dependence,
\begin{equation}
\langle x^2 \rangle = \frac{2k_BT}{m} t^2 \; \; .
\end{equation}

\noindent
The mean square displacement in the short time is proportional to $t^2$ and
is independent of the nature of the liquid. Or in other words, short
dynamics is guided by inertial motion of the particle rather than any
external influence.

\subsection{Relation between random and viscous force : The 
fluctuation-dissipation theorem}

We have the following Langevin equation
\begin{equation}
\frac{dv}{dt} = -\frac{6 \pi \eta r v}{m} + \frac{1}{m} F(t)
\end{equation}

\noindent
where, the first and the second terms are due to viscous and random forces,
respectively. We rewrite (16) after multiplying both sides by $v$
[ since $\Gamma = (6\pi \eta r)/m$ ]
\begin{equation}
v \frac{dv}{dt} = -\Gamma v^2 + \frac{1}{m} v F(t)
\end{equation}
\noindent
or
\begin{eqnarray*}
\frac{dv^2}{dt} = -2 \Gamma v^2 + \frac{2}{m} v F(t) \; \; .
\end{eqnarray*}

\noindent
Taking ensemble average we obtain
\begin{equation}
\frac{d}{dt} \langle v^2 \rangle = -2 \Gamma \langle v^2 \rangle + 
\frac{2}{m} \langle v F(t) \rangle \; \; .
\end{equation}

\noindent
The above equation requires the determination of $\langle vF(t) \rangle $;
to this end we start from the identity
\begin{equation}
\int^t_{t-\Delta t} \dot{v} (t') \; dt' =  v(t) - v(t-\Delta t)
\end{equation}
\noindent
or
\begin{equation}
v(t) =  v(t-\Delta t) + \int^t_{t-\Delta t} \dot{v} (t') \; dt' \; \; .
\end{equation}

\noindent
Multiplying both sides by $F(t)$ and taking average
\begin{equation}
\langle v(t) F(t) \rangle =  \langle v(t-\Delta t)  F(t) \rangle + 
\int^t_{t-\Delta t} \langle \dot{v} (t') F(t) \rangle \; dt' \; \; .
\end{equation}

The first term of the right hand side of Eq.(21) vanishes because of the 
velocity $v(t-\Delta t)$ at earlier instant $t-\Delta t < t$ has no
dependence on the future fluctuating force $F(t)$. Thus,
\begin{eqnarray*}
\langle v(t-\Delta t) F(t) \rangle = 0 \; \; .
\end{eqnarray*}
\noindent
Therefore
\begin{equation}
\langle v(t) F(t) \rangle =  
\int^t_{t-\Delta t} \langle \dot{v} (t') F(t) \rangle \; dt' \; \; .
\end{equation}

\noindent
Putting the expression for $\dot{v}$ in above expression
\begin{eqnarray}
\langle v(t) F(t) \rangle & = & \int^t_{t-\Delta t} \left < \left [
-\Gamma v(t') + \frac{1}{m} F(t') \right ] F(t) \right > dt' \nonumber \\
& = & \int^t_{t-\Delta t} \Gamma \langle v(t') F(t) \rangle \; dt' +
\int^t_{t-\Delta t} \frac{1}{m} \langle F(t') F(t) \rangle \; dt' \; \; ,
\end{eqnarray}

\noindent
the first term on the right hand side of Eq.(23) is zero again since 
$t'$ is the earlier time ($<t$) and the fluctuating force at a later time $t$
has no dependence on velocity earlier time $t'$ except at $t=t'$ for which the 
integral is zero. We are left with
\begin{equation}
\langle v(t) F(t) \rangle = \frac{1}{m} \int^t_{t-\Delta t}
\langle F(t') F(t) \rangle \; dt'
\end{equation}

We now assume that $F(t)$, the fluctuating force, is stationary in time. This
means the value of the integral (24) depends only on the difference $t-t'$
but not on $t$ and $t'$ individually{\footnote{In general, the stationarity
of a correlation function means that it is invariant under time translation,
i.e., $\langle F(t)F(t') \rangle = \langle F(t+T)F(t'+T) \rangle$ } }
\begin{eqnarray}
\langle v(t)F(t) \rangle & = & 
\frac{1}{m} \int^t_{t-\Delta t} \langle F(t')F(t) \rangle \; dt' \nonumber \\
& = &
\frac{1}{2m} \int^{t+\Delta t}_{t-\Delta t} \langle F(t')F(t) \rangle \; dt' \nonumber \\
& = & 
\frac{1}{2m} \int^{t+\Delta t}_{t-\Delta t} \langle F(t)F(t+s) \rangle \; ds 
\; \; , \; \; \; \; \; t=t'+s \nonumber \\
& = & \frac{1}{2m} \int^{+\infty}_{-\infty}
\langle F(0) F(s) \rangle \; ds
\end{eqnarray}

\noindent
Here it is important to note that, (i) since $\Delta t \gg (t-t'=s)$, we put
$\Delta t \sim \infty$ and (ii) The instant $t$ is long ( equilibrium time )
and arbitrary so that we may put $t=0$ without any loss of generality.

We now return to Eq.(18) and put the value of the
$\langle v(t) F(t) \rangle $ from (25) to obtain
\begin{equation}
\frac{d}{dt} \langle v^2 \rangle = -2\Gamma \langle v^2 \rangle +
\frac{1}{m} \int^{+\infty}_{-\infty} ds  \; \langle F(0) F(s) \rangle
\end{equation}

\noindent
In thermal equilibrium, the time derivative vanishes and the law of
equipartition states
\begin{equation}
\frac{1}{2} m \langle v^2 \rangle = \frac{1}{2} k_BT
\end{equation}

\noindent
Eq.(26) therefore gives,
\begin{eqnarray*}
0 = -2\Gamma \langle v^2 \rangle +
\frac{1}{m} \int^{+\infty}_{-\infty} ds  \; \langle F(0) F(s) \rangle
\end{eqnarray*}

\noindent
Applying (27) we obtain
\begin{equation}
\Gamma = \frac{1}{2m k_BT} \int^{+\infty}_{-\infty} ds \; 
\langle F(0) F(s) \rangle \; \; .
\end{equation}

The above equation is known as the fluctuation-dissipation theorem, since
it relates the dissipation $\Gamma$ to the correlation of the fluctuating force
$F(t)$. It expresses a balance between the input of energy flow into the
system ( particle ) due to the fluctuating forces imparted by the liquid
and the output of energy flow from the system due to dissipative forces
exerted by the liquid on the system.

One can easily have a relation between the diffusion in velocity space and
viscosity. For this purpose we require that the fluctuating forces $F(t)$
are instantaneously correlated, i.e., we use
\begin{eqnarray*}
\langle F(0) F(s) \rangle = 2{\cal D} \delta(s) \; \; .
\end{eqnarray*}

\noindent
Also since, $\Gamma = (6\pi \eta r/m)$, we have from Eq.(28)
\begin{eqnarray*}
\frac{6\pi \eta r}{m} = \frac{ {\cal D} }{m k_BT}
\int^{+\infty}_{-\infty} \delta(s) ds
\end{eqnarray*}

\noindent
or
\begin{eqnarray*}
{\cal D} = 6 \pi \eta r k_B T \; \; .
\end{eqnarray*}

\noindent
Thus this relation may be visualized as special form of
fluctuation-dissipation theorem.

\section{Brownian motion in velocity space : Fokker-Planck equation}

\setcounter{equation}{0}

\subsection{Fokker-Planck equation}

In Einstein's method we considered the problem of Brownian motion in co-ordinate 
space, i.e., it concerns the time development of distribution of suspended
particles in terms of $f(x,t)$, or the probability of finding a particle at
the position $x$ at a time $t$. The law of evolution was stated to be,
\begin{equation}
f(x, t+\tau) = \int^{+\infty}_{-\infty} f(x+\Delta, t) \; \phi(\Delta)
\; d\Delta
\end{equation}

\noindent
where one takes into account of all the possible jumps of magnitude $\Delta$
from $x+\Delta$ to $x$, each with probability $\phi(\Delta)$. The differential 
form of the above equation is the diffusion equation,
\begin{equation}
\frac{\partial f(x,t)}{\partial t} = D \frac{\partial^2 f(x,t)}{\partial x^2}
\end{equation}

\noindent
where $D = \int^{+\infty}_{-\infty} \frac{\Delta^2}{2} \phi(\Delta) \; d\Delta$
is the diffusion coefficient in co-ordinate space.

Herein we approach the problem of Brownian motion in velocity space (as in
Langevin description ) and are concerned with the probability $f(v,t)$ that
a particle has a velocity $v$ at a time $t$. The technique is applicable to 
any fluctuating quantity. For the sake of simplicity we consider, however, 
the problem in one dimension.

The time development of probability distribution $f(v,t)$ of velocities may 
be stated as,
\begin{equation}
f(v,t+\tau) = \int^{+\infty}_{-\infty} f(v-\Delta, t) \;
\phi(v-\Delta, \Delta) \; d\Delta
\end{equation}

\noindent
Here $\phi(v-\Delta, \Delta)$ is the probability of a jump $\Delta$ for a
particle with velocity $v-\Delta$. Note that $\Delta$ has a dimension of 
velocity. Both $\tau$ and $\Delta$ are small such that higher power of them 
may be neglected in the calculation. We now expand $f(v,t+\tau)$ around $t$ 
and $f(v-\Delta,t) \; \phi(v-\Delta,\Delta)$ around $v$ such that
\begin{equation}
f(v,t+\tau) = f(v,t) + \tau \frac{\partial f}{\partial t} + \frac{\tau^2}{2}
\frac{\partial^2 f}{\partial t^2} + \ldots
\end{equation}
\begin{equation}
f(v-\Delta,t) \phi(v-\Delta, \Delta) = f(v,t) \phi(v,\Delta) - \Delta
\frac{\partial (f\phi)}{\partial v} + \frac{\Delta^2}{2}
\frac{\partial^2 (f\phi)}{\partial v^2}
\end{equation}

\noindent
Putting (4) and (5) in (3) we obtain,
\begin{eqnarray}
f(v,t) + \tau \frac{\partial f}{\partial t} + {\cal O}(\tau^2)
& = & \int^{+\infty}_{-\infty} \; \left [ f(v,t) \phi(v,\Delta) - \Delta
\frac{\partial (f\phi)}{\partial v} + \frac{\Delta^2}{2}
\frac{\partial^2 (f\phi)}{\partial v^2} \right ] \; d\Delta \nonumber \\
& = & f(v,t) \int^{+\infty}_{-\infty} \phi(v,\Delta) \; d\Delta -
\frac{\partial}{\partial v} \left [ f(v,t) \int^{+\infty}_{-\infty}
\Delta \; \phi(v,\Delta) \; d\Delta \right ]  \nonumber \\
& & + \frac{\partial^2}{\partial v^2} \left [ f(v,t) \int^{+\infty}_{-\infty}
\frac{\Delta^2}{2} \phi(\Delta, v) \; d\Delta \right ]
\end{eqnarray}

\noindent
Note that $\int^{+\infty}_{-\infty} \phi(v,\Delta) \; d\Delta = 1$
[ probability is normalized ]. We then write,
\begin{eqnarray*}
\frac{1}{\tau} \; \int^{+\infty}_{-\infty} \Delta \; \phi(v,\Delta) \;
d \Delta & = & \frac{ \langle \Delta (v) \rangle}{\tau} \; \; \; =
\; \; \; M_1(v) \nonumber \\
\frac{1}{\tau} \; \int^{+\infty}_{-\infty} \frac{\Delta^2}{2} 
\; \phi(v,\Delta) \; d\Delta & = & 
\frac{ \langle \Delta^2 (v) \rangle}{\tau} \; \; \; = \; \; \; M_2(v) 
\end{eqnarray*}

\noindent
Eq.(6) can be rewritten as,
\begin{equation}
\frac{\partial f(v,t)}{\partial t} = - \frac{\partial}{\partial v} M_1(v) \;
f(v,t) + \frac{\partial^2}{\partial v^2} M_2(v) \; f(v,t)
\end{equation}

\noindent
$M_1(v)$ and $M_2(v)$ are called the drift and diffusion terms, respectively.
Eq.(7) is called the Fokker-Planck equation. $M_2$ is the diffusion in 
velocity space and is not to be confused with$D$ of Eq.(2) in co-ordinate space.

For complete specification of Fokker-Planck equation one therefore needs the
information about the drift, $M_1(v)$ and diffusion, $M_2(v)$ terms in Eq.(7).
We calculate here these quantities for a specific model, i.e., the Brownian motion.

\subsection{Calculation of $M_1(v)$}

We integrate the Langevin equation
\begin{equation}
\frac{dv}{dt} = -\alpha v + F(t)
\end{equation}

\noindent
between $t$ and $t+\tau$ to obtain
\begin{equation}
v(t+\tau) -v(t) = -\alpha v \tau + \int_t^{t+\tau} F(t') \; dt'
\end{equation}

\noindent
$\alpha$ in Eq.(8) is the dissipation or damping constant and $F(t)$ is the
random force whose average is zero, i.e., $\langle F(t) \rangle = 0$. We now 
put $\Delta = v(t+\tau)-v(t)$, therefore,
\begin{eqnarray*}
\langle v(t+\tau) -v(t) \rangle = \langle \Delta \rangle = -\alpha v \tau
\end{eqnarray*}

\noindent
or
\begin{equation}
M_1(v) = \frac{ \langle \Delta \rangle }{\tau} = -\alpha v
\end{equation}

\noindent
is the drift term.

\subsection{Calculation of $M_2(v)$}

From Eq.(9) we write
\begin{equation}
\Delta^2 = \alpha^2 v^2 \tau^2 - 2\alpha v \tau \int^{t+\tau}_t F(t') \; dt'
+ G_t^2(\tau)
\end{equation}

\noindent
where $G_t(\tau) = \int^{t+\tau}_t F(t') \; dt'$. Since $\tau$ is very small, 
the first term is neglected. Also the second term by virtue of 
$\langle F(t) \rangle = 0$ is zero. So we are left with ( after averaging )
\begin{eqnarray*}
\langle \Delta^2 \rangle = \langle G_t^2 (\tau) \rangle
\end{eqnarray*}
\begin{equation}
\langle \Delta^2 \rangle = \int^{t+\tau}_t dt' \int^{t+\tau}_t
\langle F(t') F(t'') \rangle \; dt''
\end{equation}

\noindent
If we assume that the random force is instantaneously correlated, i.e.,
\begin{eqnarray*}
\langle F(t') F(t'') \rangle = 2 {\cal D} \delta(t'-t'')
\end{eqnarray*}

\noindent
we obtain the average of $\Delta^2$ as
\begin{eqnarray*}
\langle \Delta^2 \rangle & = & 2 {\cal D} \int^{t+\tau}_t dt' \int^{t+\tau}_t
\delta(t'-t'') \; dt'' \nonumber \\
& = & 2 {\cal D} \int^{t+\tau}_t dt'
\end{eqnarray*}

\noindent
or
\begin{equation}
\langle \Delta^2 \rangle = 2 {\cal D} \tau
\end{equation}

\noindent
Therefore, $M_2(v) = \frac{ \langle \Delta^2 \rangle }{2\tau} = {\cal D}$,
the diffusion constant in velocity space, which is to be determined for
the model represented by Eq.(8).

To determine ${\cal D}$ we first integrate Eq.(8) formally with the 
integrating factor $e^{-\alpha t}$ and obtain
\begin{equation}
v(t) = v(0) e^{-\alpha t} + e^{-\alpha t} \int^t_0 e^{-\alpha t'} F(t') \; dt'
\end{equation}

\noindent
Rearranging and taking the square on both sides we have
\begin{equation}
( v(t) -v(0) e^{-\alpha t} )^2 = e^{-2\alpha t} \int_0^t dt' \int_0^t
e^{\alpha(t'+t'')} F(t') F(t'') \; dt''
\end{equation}

\noindent
Averaging yields
\begin{equation}
\langle ( v(t) -v(0) e^{-\alpha t} )^2 \rangle = e^{-2\alpha t} \int_0^t dt' 
\int_0^t e^{\alpha(t'+t'')} \langle F(t') F(t'') \rangle \; dt''
\end{equation}

\noindent
Putting $\langle F(t') F(t'') \rangle = 2{\cal D} \delta(t'-t'')$ 
in (16) we obtain
\begin{equation}
\langle ( v(t) -v(0) e^{-\alpha t} )^2 \rangle = 2 e^{-2\alpha t} {\cal D}
\int^t_0 dt' e^{2\alpha t'}
\end{equation}

\noindent
Explicit integration in the last equation yields
\begin{eqnarray*}
\langle ( v(t) -v(0) e^{-\alpha t} )^2 \rangle & = & 2 e^{-2\alpha t} {\cal D}
\left [ \frac{ e^{2\alpha t} -1}{2\alpha} \right ] \nonumber\\
& = & \frac{ {\cal D}}{\alpha} ( 1-e^{-2\alpha t} )
\end{eqnarray*}

\noindent
For $t \gg 1/\alpha$, we have
\begin{equation}
\langle v(t)^2 \rangle = \frac{ {\cal D}}{\alpha} \; \; .
\end{equation}

The above relation together with the equipartition theorem
$\frac{1}{2}m\langle v^2 \rangle = \frac{1}{2} k_B T$ leads us to
\begin{equation}
{\cal D} = \alpha k_B T / m
\end{equation}

\noindent
Thus $M_2(v) = {\cal D} = \alpha k_B T/m$, ( diffusion coefficient 
in velocity space ).

The above ${\cal D}$ is not to be confused $D$ of Eq.(22) which is a diffusion
coefficient for the Brownian particle in co-ordinate space.

With these $M_1$ and $M_2$ the Fokker-Planck equation for the Brownian motion can be 
rewritten as
\begin{equation}
\frac{\partial f(v,t)}{\partial t} = \frac{\partial}{\partial v} [\alpha v ]
f(v,t) + \frac{\partial^2}{\partial v^2} \left [ 
\frac{\alpha k_B T}{m} \right ] f(v,t)
\end{equation}

\noindent
Putting $p=mv$ the above equation (20) can be rewritten in the momentum
space as follows
\begin{equation}
\frac{\partial P(p,t)}{\partial t} = \frac{\partial}{\partial p} [\alpha p ]
P(p,t) + \frac{\partial^2}{\partial p^2} \left [ 
\frac{\alpha k_B T}{m} \right ] P(p,t)
\end{equation}

\noindent
where $f(v,t) \equiv P(p,t)$ represents the probability distribution function
in momentum space. Here the underlying stochastic process is called the 
Ornstein-Uhlenbeck process (1930).

\section{
Brownian motion in phase space (motion in a force field) }

\setcounter{equation}{0}

\subsection{Kramers' equation}

We first derive here the equation of diffusion for an ensemble of particles with
probability density distribution $f(p, q, t)$ in phase space (i.e., q, p-space).
The evolution of distribution from the time $t$ to another time $t+\tau$ is 
given by the following equation,
\begin{equation}
f(p_1, q_1, t+\tau) = \int_{-\infty}^{+\infty} f(p-\Delta, q, t) 
\phi(p-\Delta, q, \Delta) d\Delta
\end{equation}

Had there been no Brownian motion, the motion would have been purely
deterministic, i.e.,
\begin{equation}
\left. \begin{array}{ccc}
\dot{q} & = & p ,\\
\dot{p} & = & {\cal K}(q) \end{array} \right \}
\end{equation}

\noindent
where ${\cal K}(q)$ is the force acting on the particles. Thus the time 
development of $q$ and $p$ over a small time $\tau$ would be
\begin{equation}
q_1 = q + p\tau \; \; {\rm and} \; \; p_1 = p + {\cal K}\tau \; \; .
\end{equation}

\noindent
Here $(q, p)$ and $(q_1, p_1)$ specify the co-ordinate-momentum pair at time 
$t$ and $t+\tau$.

Because the particle is also subjected to a random Brownian force, we account
for the all the possible jumps (of magnitude $\Delta$) in momentum with the
probability function $\phi(p-\Delta, q, \Delta)$ and an integration over 
$\Delta$ in Eq.(1).

Making use of Eq.(3) we rewrite Eq.(1) as,
\begin{equation}
f(p+{\cal K}\tau, q+p\tau, t+\tau) = \int_{-\infty}^{+\infty} f(p-\Delta, q, t)
\phi(p-\Delta, q, \Delta) d\Delta
\end{equation}

We now expand $f(p+{\cal K}\tau, q+p\tau, t+\tau)$ in a Taylor series as,
\begin{equation}
f(p, q, t) + \frac{\partial f}{\partial p} {\cal K}\tau + \frac{\partial f}
{\partial q} p\tau + \frac{\partial f}{\partial t}\tau + \ldots
\end{equation}

Also expanding $f(p-\Delta, q, t) \phi(p-\Delta, q, \Delta)$ as,
\begin{equation}
f(p, q, t)\phi(p, q, \Delta) - \frac{\partial (f\phi)}{\partial p} \Delta
+ \frac{1}{2!} \frac{\partial^2 (f\phi)}{\partial p^2} \Delta^2 + \ldots
\end{equation}

Integration over $\Delta$ in Eq.(6) gives
\begin{eqnarray*}
\int_{-\infty}^{+\infty} f(p-\Delta, q, t) \phi(p-\Delta, q, \Delta) d\Delta
\end{eqnarray*}
\begin{eqnarray*}
= f \int_{-\infty}^{+\infty} \phi(p, q, \Delta) d\Delta
-\frac{\partial}{\partial p} [ f ] \int_{-\infty}^{+\infty} \Delta 
\phi(p, q, \Delta) d\Delta
+\frac{\partial^2}{\partial p^2} [ f ] \int_{-\infty}^{+\infty} 
\frac{\Delta^2}{2} \phi(p, q, \Delta) d\Delta 
\end{eqnarray*}
\begin{equation}
= f(p, q, t) - \frac{\partial}{\partial p} [ f(p, q, t) M_1(p, q) ] +
\frac{\partial^2}{\partial p^2} [ f(p, q, t) M_2(p, q) ] \; \; .
\end{equation}

\noindent
We now put
\begin{eqnarray*}
\begin{array}{ccc}
\int_{-\infty}^{+\infty} \phi(p, q, \Delta) d\Delta & = & 1 \\
\nonumber \\
\int_{-\infty}^{+\infty} \Delta \phi(p, q, \Delta) d\Delta & = & 
{\overline \Delta} \\
\nonumber \\
\int_{-\infty}^{+\infty} \frac{\Delta^2}{2} 
\phi(p, q, \Delta) d\Delta & = & {\overline{\Delta^2}}/2
\end{array}
\end{eqnarray*}

Using Eq.(5) and Eq.(7) in Eq.(4) we obtain
\begin{equation}
\frac{\partial f}{\partial p} {\cal K}\tau + \frac{\partial f}{\partial q} p\tau
+ \frac{\partial f}{\partial t}\tau = 
-\frac{\partial}{\partial p} [ f {\overline \Delta}] 
+ \frac{\partial^2}{\partial p^2} [ f {\overline {\Delta^2}} ]
\end{equation}

Dividing both sides of Eq.(8) by $\tau$ we get
\begin{equation}
\frac{\partial f}{\partial p} {\cal K} + \frac{\partial f}{\partial q} +
\frac{\partial f}{\partial t} = -\frac{\partial}{\partial p}(f M_1) +
\frac{\partial^2}{\partial p^2} (f M_2)
\end{equation}

\noindent
where $M_1 = {\overline {\Delta}}/\tau$ and 
$M_2 = {\overline {\Delta^2}}/2\tau$. Thus rearranging Eq.(9) we write
\begin{equation}
\frac{\partial f}{\partial t} = -p\frac{\partial f}{\partial q} -
{\cal K}(q)\frac{\partial f}{\partial p} - \frac{\partial}{\partial p}
[ M_1(p, q) f ] + \frac{\partial^2}{\partial p^2} [ M_2(p, q) f ] \; \; .
\end{equation}

Since the force ${\cal K}(q)$ is derivable from a potential $V(q)$ we write
\begin{equation}
{\cal K}(q) = -V'(q)
\end{equation}

From the knowledge of Brownian motion (as calculated in the last section)
we know,
\begin{equation}
\left. \begin{array}{ccccc}
M_1 & = & -\gamma p & = & -m\gamma v \\
M_2 & = & {\cal  D} & = & m\gamma k_B T \end{array} \right \}
\end{equation}

Take the mass of the particle $m=1$ for simplicity. Then writing $q=x$ and 
$p=v$ Eq.(10) reduces to
\begin{equation}
\frac{\partial f}{\partial t} = -\frac{\partial f}{\partial x} v +
\frac{\partial f}{\partial v} [V'(x)] + \frac{\partial}{\partial v}
[\gamma v]f + \gamma k_BT \frac{\partial^2 f}{\partial v^2}
\end{equation}

The above equation is called the Kramers' equation \cite{kramers,rmp3}. 
It describes the 
Brownian motion of a particle which is in a field of force. Or in other words
a particle moves in an external field but in addition is subjected to irregular
forces (at the same time) of the surrounding medium. While the first two terms
are due to the deterministic motion, the third and the fourth terms are the 
drift and diffusion terms which are characteristic of Brownian motion. 

\subsection{Kramers equation as a generalization of Liouville equation and
connection to equilibrium statistical mechanics}

The deterministic motion described by (13), i.e., 
\begin{equation}
\frac{\partial f(x,v,t)}{\partial t} = -\frac{\partial f}{\partial x} v +
\frac{\partial f}{\partial v} [V'(x)] 
\end{equation}

\noindent
corresponds to Liouville equation which forms the basis of equilibrium 
statistical mechanics under the condition $\frac{\partial f}{\partial t}=0$ 
( which defines the equilibrium )

It is now wellknown that under equilibrium condition the distribution is a 
Maxwell-Boltzmann distribution, i.e., ( we assume $m=1$ )
\begin{equation}
f(x,v) = Z \; e^{- \frac{ \frac{1}{2} v^2 + V(x) }{ k_B T} }
\end{equation}

\noindent
where $Z$ is the normalization constant. Thus (15) satisfies (14) for
$\frac{\partial f}{\partial t}=0$  as may be checked.

It is important to emphasize that if we keep the Brownian dynamical terms 
in (13) as such and put the equilibrium condition
$\frac{\partial f}{\partial t}=0$  then it may also be checked that (15) 
satisfies (13) under the equilibrium condition 
$\frac{\partial f}{\partial t}=0$ .

Kramers' equation may thus be regarded as a generalization of Liouville
equation since it includes the Brownian motion in such a way that the basis 
of equilibrium statistical mechanics remains unaffected.

\subsection{Kramers' theory of activated processes}

\setcounter{equation}{0}

Kramers' model for a chemical reaction consists of a classical particle of mass 
$m$ (considered here to be unity) moving in a one-dimensional asymmetric
double-well potential $V(x)$. The particles co-ordinate $x$ corresponds to the 
reaction co-ordinate and its value at the minima of the potential $V(x)$,
$x_a$ and $x_c$ denotes the reactant and the product states, respectively. The
maximum of $V(x)$ at $x=x_b$ separating these states corresponds to the 
transition state (or activated complex). All the remaining degrees of freedom
of reactants and the solvent molecules constitute the surrounding medium whose
total effect on the reacting particle is described by a fluctuating force and 
a linear damping. The correlation of fluctuating force gives rise to diffusion
coefficient and the linear damping is responsible for the drift term. The 
stochastic dynamics for the reaction co-ordinate $x$ and velocity $v$ is
governed by Kramers' equation
{\footnote {here we have put $f(x, v, t) = P(x, v, t)$ ;
just a change in notation}}
\begin{equation}
\frac{\partial}{\partial t} P(x, v, t) = \left [ -\frac{\partial}{\partial x}
v + \frac{\partial}{\partial v} \left \{ V'(x) +\gamma v \right \} +
\gamma k_B T \frac{\partial^2}{\partial v^2} \right ] P(x, v, t)
\end{equation}

The conditions are such that the particle is originally caught in the left well
may escape in the course of time due to thermal activation 
by passing over the potential barrier. We
want to calculate the probability of escape and its dependency on temperature
and viscosity of the medium and compare the value with the result of `transition
state method'. The calculation rests on the equation of diffusion obeyed by a
density-distribution of particles in phase space as written above [Eq. (1)].

To determine the steady state escape rate from A to C (say) we consider that
there is a stationary situation in which a steady state probability current
(flux) over B from A$\rightarrow$C is maintained.

The stationary probability density must satisfy the following conditions :

\noindent
$\bullet$ Since we are considering a stationary situation, i.e., 
$\frac{\partial P}{\partial t}=0$ we write,
\begin{equation}
\left [ -\frac{\partial}{\partial x} v + \frac{\partial}{\partial v}
\left \{ V'(x) + \gamma v \right \} + \gamma k_B T 
\frac{\partial^2}{\partial v^2} \right ] P(x, v) = 0 
\end{equation}

\noindent
$\bullet\bullet$ At the barrier top B we assume the linearized potential, 
i.e., we write by expanding $V(x)$ around $x_b$
\begin{eqnarray*}
V(x) = V(x_b) + \left. \frac{\partial V}{\partial x} \right |_{x=x_b}
(x-x_b) + \frac{1}{2} \left. \frac{\partial^2 V}{\partial x^2}
\right |_{x=x_b} (x-x_b)^2
\end{eqnarray*}

Since $\left. \frac{\partial V}{\partial x} \right |_{x=x_b} = 0$
and $-\left. \frac{\partial^2 V}{\partial x^2} \right |_{x=x_b} = \omega_b^2$
we have,
\begin{equation}
V(x) = V(x_b) - \frac{1}{2}\omega_b^2 (x-x_b)^2 \; \; .
\end{equation}

While considering the motion around $x=x_b$ the above potential (3) has to be 
used in Eq.(2).

\noindent
$\bullet\bullet\bullet$ Near the bottom of the A-well all the particles 
are thermalized. Therefore we must have the usual 
Boltzmann distribution to be valid here, i.e. ,
\begin{equation}
P(x, v) = z^{-1} \exp\left [ \left \{ -\frac{1}{2}v^2 + V(x) \right \}
/k_B T \right ] \; \; {\rm at} \; x \approx x_a
\end{equation}

The linearization of potential has to be done at $x=x_a$, i.e., we write
\begin{eqnarray*}
V(x) = V(x_a) + \left. \frac{\partial V}{\partial x} \right |_{x=x_a}
(x-x_a) + \frac{1}{2} \left. \frac{\partial^2 V}{\partial x^2}
\right |_{x=x_a} (x-x_a)^2
\end{eqnarray*}

Since $\left. \frac{\partial V}{\partial x} \right |_{x=x_a} = 0$
and $-\left. \frac{\partial^2 V}{\partial x^2} \right |_{x=x_a} = \omega_a^2$
we have,
\begin{equation}
V(x) = V(x_a) + \frac{1}{2}\omega_a^2 (x-x_a)^2 \; \; .
\end{equation}

While considering the motion around $x=x_a$, the potential $V(x)$ as given 
by Eq.(5) has to be used in Eq.(2)

\noindent
$\bullet\bullet\bullet\bullet$ Near the bottom of the well C all the 
particles are (as if) removed. This implies the condition
\begin{equation}
P(x, v) \sim 0 \; \; \; {\rm for} \; x > x_b \;\; .
\end{equation}

Once the probability density $P(x, v)$ fulfilling the above requirements
is known, the population in the A-well $n_a$ and the flux $j$ over the barrier
will be given by
\begin{eqnarray}
n_a = \int_{\rm A well} dx \; dv \; P(x, v) \\
j = \int^{+\infty}_{-\infty} dv \; v \; P(x, v)
\end{eqnarray}

Hence the steady state Kramers' rate is given by
\begin{equation}
k_{A \rightarrow C} = j/n_a \; \; .
\end{equation}

Our next task is to calculate $j$ and $ n_a$ separately.

\subsubsection{Calculation of j}

Since we are considering the flux over the barrier B, the linearized potential
$V(x) = V(x_b) -\frac{1}{2} \omega_b^2 (x-x_b)^2$ has to be used. The Kramers'
equation{\footnote {since $V'(x) =  -\omega_b^2(x-x_b)$ }} therefore reduces to
\begin{equation}
\left [ -v\frac{\partial}{\partial x} + \frac{\partial}{\partial v} \left \{
-\omega_b^2(x-x_b) + \gamma v \right \} + \gamma k_B T \frac{\partial^2}
{\partial v^2} \right ] P(x,v) = 0 \; \; .
\end{equation}

We now construct $P(x,v)$ in the following form
\begin{equation}
P(x,v) = \xi(x,v) \exp \left [ \frac{-\frac{1}{2} v^2 + V(x) }{k_B T} \right ]
\; \; , \; \; x\approx x_b \; \; .
\end{equation}

Putting Eq.(11) in Eq.(10) we get after a little bit of straightforward
algebra
\begin{equation}
\left [ -v\frac{\partial}{\partial x} - \left \{ \omega^2_b (x-x_b) +
\gamma v \right \} \frac{\partial}{\partial v} + \gamma k_BT 
\frac{\partial^2}{\partial v^2} \right ] \xi (x,v) =0 \; \; .
\end{equation}

Boundary condition for $\xi(x,v)$ in Eq.(11) should be such that

\noindent
$\bullet$ $\xi(x,v) \rightarrow 1$ inside the well A, for $x\approx x_a$.

\noindent
$\bullet$ $\xi(x,v) \rightarrow 0$ beyond the barrier top B, for $x > x_b$.

We now use the following linear transformation
\begin{equation}
u = v \; + \; a(x-x_b)
\end{equation}

\noindent
where $a$ is a constant to be determined later. This gives
\begin{equation}
\left. \begin{array}{ccc}
\frac{\partial}{\partial x} & = & a \frac{\partial}{\partial u} \\
\frac{\partial}{\partial v} & = & \frac{\partial}{\partial u} 
\end{array} \right \}
\end{equation}

Making use of Eq.(13) and Eq.(14) we obtain from Eq.(12)
\begin{equation}
-av \frac{\partial \xi}{\partial u} - \left [ \omega_b^2(x-x_b) + \gamma v
\right ] \frac{\partial \xi}{\partial u} + \gamma k_BT
\frac{\partial^2 \xi}{\partial u^2} = 0
\end{equation}
\noindent
or
\begin{equation}
\gamma k_B T\frac{\partial^2 \xi}{\partial u^2} - \left [ \omega_b^2 (x-x_b)
+ v(a+\gamma) \right ]\frac{\partial \xi}{\partial u} = 0
\end{equation}

We now put :
\begin{eqnarray*}
\omega_b^2(x-x_b) + v(a+\gamma) = -\lambda u \; \; , \; \; \lambda =
{\rm constant \; (to \; be \; determined)}.
\end{eqnarray*}
\noindent
Since
\begin{eqnarray*}
u=v \; + \; a(x-x_b)
\end{eqnarray*}
\noindent
we have,
\begin{eqnarray*}
\omega_b^2 (x-x_b) \; + \; v(a+\gamma) = -\lambda [ v+a(x-x_b) ] \; \; .
\end{eqnarray*}

Comparing both sides
\begin{equation}
\left. \begin{array}{ccc}
-\lambda a & = & \omega_b^2 \\
-\lambda  & = & a+\gamma
\end{array}\right \}
\end{equation}

From Eq.(17) we eliminate $\lambda$ to obtain
\begin{eqnarray*}
a^2 + \gamma a - \omega_b^2 = 0
\end{eqnarray*}
\noindent 
which gives
\begin{equation}
a=-\frac{\gamma}{2} \pm \sqrt{ \left ( \frac{\gamma}{2} \right )^2 +
\omega_b^2 }
\end{equation}
\noindent
and
\begin{eqnarray*}
\lambda = -\omega_b^2 / a \; \; .
\end{eqnarray*}

Eq.(16) now reduces to
\begin{equation}
\gamma k_BT \frac{\partial^2 \xi}{\partial u^2} + \lambda u 
\frac{\partial \xi}{\partial u} = 0
\end{equation}
\noindent
or
\begin{equation}
\frac{\partial^2 \xi}{\partial u^2} + \frac{\lambda u}{\gamma k_BT}
\frac{\partial \xi}{\partial u} = 0
\end{equation}

Our next task is to solve Eq.(20). We put
$\frac{\partial \xi}{\partial u} = y$. Therefore Eq.(20) reduces to
\begin{equation}
\frac{\partial y}{\partial u} = -\frac{\lambda}{\gamma k_BT} u y
\end{equation}

Integrating over $u$ Eq.(21) gives
\begin{eqnarray*}
\ln y = -\frac{\lambda}{\gamma k_BT} u^2 + \ln F_2  \; \; , \; \;
F_2 = {\rm constant \; of \; integration}
\end{eqnarray*}
\begin{eqnarray*}
y = F_2 \; \exp\left (-\frac{\lambda u^2}{2 \gamma k_BT} \right )
\end{eqnarray*}

Since
\begin{eqnarray*}
\frac{\partial \xi}{\partial u} =
y = F_2 \; \exp\left (-\frac{\lambda u^2}{2 \gamma k_BT} \right )
\end{eqnarray*}
\noindent
we get
\begin{equation}
\xi(u) = F_2 \int_0^u \exp \left ( -\frac{\lambda u^2}{2\gamma k_BT} \right )
+ F_1 \; \; , \; \; F_1 = {\rm constant \; of \; integration}
\end{equation}

We look for a solution that vanishes at $x\rightarrow \infty $; the above 
integral should however remain finite for all $|u| \rightarrow \infty$.
This implies $\lambda > 0$, i.e., positive. Since
\begin{equation}
\lambda = - \omega_b^2 / a
\end{equation}

\noindent
the negative root of $a$ should be chosen to keep $\lambda$ positive, i.e.,
\begin{equation}
a = -\frac{\gamma}{2} - \sqrt{\left ( \frac{\gamma}{2} \right )^2 +
\omega_b^2 } \; \; .
\end{equation}

Thus $\lambda$ and $a$ are known in terms of the given parameters $\gamma$
and $\omega_b^2$ of the problem.

Next we determine $F_1$ and $F_2$ (the integration constants).

When $x\rightarrow \infty$ then $u \rightarrow -\infty$, since $u=v+a(x-x_b)$
and $a$ is negative. Again when $x \rightarrow \infty$ we must have 
$\xi(x,v) \rightarrow 0$. Therefore from Eq.(22) we obtain
\begin{eqnarray*}
0 =F_1 + F_2 \int_0^{-\infty} \exp \left ( -\frac{\lambda u^2}{2\gamma k_BT}
\right ) \; du
\end{eqnarray*}
\noindent
or
\begin{eqnarray}
F_1 & = & F_2 \int_{-\infty}^0 \exp \left (-\frac{\lambda u^2}{2 \gamma k_BT}
\right ) \; du \nonumber \\
F_1 & = & F_2 \; \frac{1}{2} \; \int_{-\infty}^{+\infty} 
\exp \left (-\frac{\lambda u^2}{2 \gamma k_BT} \right ) \; du \nonumber \\
F_1 & = & F_2 \sqrt{\frac{\pi \gamma k_BT}{2 \lambda}}
\end{eqnarray}

Therefore we obtain
\begin{equation}
\xi(u) = F_2 \left [ \sqrt{\frac{\pi \gamma k_BT}{2\lambda}} + \int_0^u
\exp \left (- \frac{\lambda u^2}{2\gamma k_BT} \right ) \; du \right ] \; \; .
\end{equation}

We then return to the expression for $P(x,v)$
\begin{equation}
P(x,v) = \xi (x,v) \exp \left [ - \frac{\frac{1}{2} v^2 + V(x_b) -
\frac{1}{2} \omega_b^2 (x-x_b)^2}{k_BT} \right ] \; , \; x\approx x_b \; \; .
\end{equation}

Using Eq.(26) we get from Eq.(27)
\begin{eqnarray}
P(x,v) = F_2 \; \left [ \sqrt{\frac{\pi \gamma k_BT}{2\lambda}} + \int_0^u
\exp \left (-\frac{\lambda u^2}{2\gamma k_BT} \right ) \; du \right ] 
\nonumber \\
\times \exp \left [ -\frac{ \frac{1}{2}v^2 + V(x_b) -\frac{1}{2} \omega_b^2
(x-x_b)^2 }{k_BT} \right ] \; \; .
\end{eqnarray}

Writing
\begin{eqnarray*}
C= \sqrt{\frac{\pi \gamma k_BT}{2\lambda} }
\end{eqnarray*}
\noindent
and
\begin{eqnarray*}
F(x,v) = \int_0^u \exp\left (-\frac{\lambda u^2}{2\gamma k_BT} \right ) \; du
\end{eqnarray*}
\noindent
Eq.(28) reduces to
\begin{eqnarray}
P(x_b,v) = F_2 \; \exp \left (-\frac{V(x_b)}{k_BT} \right ) \nonumber \\
\times \left [
C \; \exp \left ( -\frac{1}{2} v^2 /k_BT \right ) \; + \; F(x_b, v) \;
\exp \left ( -\frac{1}{2} v^2 /k_BT \right ) \right ] \; , \; x\approx x_b
\end{eqnarray}

The expression for the current $j$ is given by
\begin{equation}
j = \int^{+\infty}_{-\infty} v \; P(x_b,v) \; dv
\end{equation}

Since the first term in Eq.(29) can not contribute to Eq.(30) because of
\begin{eqnarray*}
\int^{+\infty}_{-\infty} v e^{-\frac{1}{2}v^2/k_BT} dv =0
\end{eqnarray*}

\noindent
we write
\begin{eqnarray}
j & = & F_2 \; e^{-\frac{V(x_b)}{k_BT} } \left [ \int_{-\infty}^{+\infty}
v \; e^{-v^2/2k_BT} \; F(x_b,v) \; dv \right ] \nonumber \\
& = & F_2 \; e^{-\frac{V(x_b)}{k_BT} } \; (-k_BT) \int_{-\infty}^{+\infty}
\frac{\partial}{\partial v} \; e^{-v^2/2k_BT} \; F(x_b,v) \; dv
\nonumber \\
& = & F_2 \; e^{-\frac{V(x_b)}{k_BT} } \; (-k_BT) \; \left \{ \left. F(x,v)
e^{-v^2/2k_BT} \right |_{-\infty}^{+\infty} - \int_{-\infty}^{+\infty}
\frac{\partial F}{\partial v} \; e^{-v^2/2k_BT} \; dv \right \}
\nonumber \\
& = & F_2 \; e^{-\frac{V(x_b)}{k_BT} } \; (k_BT) \; \int_{-\infty}^{+\infty}
\frac{\partial F}{\partial v} \; e^{-v^2/2k_BT} \; dv
\end{eqnarray}

Since{\footnote {since at $x=x_b$  $u=v$} }
\begin{eqnarray*}
F(x,v) & = & \int_0^u \exp \left (-\frac{\lambda u^2}{2\gamma k_BT} \right )
\; du \nonumber \\
\frac{\partial F}{\partial v} & = & \exp \left (- \frac{\lambda v^2}
{2\gamma k_BT} \right ) \; \; .
\end{eqnarray*}

Therefore Eq.(28) gives
\begin{eqnarray}
j & = & F_2 \; e^{-\frac{V(x_b)}{k_BT} } \; k_BT \; \int_{-\infty}^{+\infty}
e^{-\frac{\lambda v^2}{2\gamma k_BT} } \; e^{-\frac{v^2}{2k_BT} } \; dv \\
& = & F_2 \; e^{-\frac{V(x_b)}{k_BT} } \; k_BT \; \int_{-\infty}^{+\infty} \;
e^{- \left [ \frac{\lambda}{2\gamma k_BT} + \frac{1}{2k_BT} \right ] v^2 }
\; dv \\
& = & F_2 \; e^{-\frac{V(x_b)}{k_BT} } \; (k_BT) \; (2\pi k_BT)^{1/2} \;
\left ( \frac{\gamma}{\lambda + \gamma} \right )^{1/2} \; \; .
\end{eqnarray}

Finally we get the expression for steady state current
\begin{equation}
j = F_2 \; e^{-\frac{V(x_b)}{k_BT} } \; (2\pi)^{1/2} \; (k_BT)^{3/2} \;
\left ( \frac{\gamma}{\lambda +\gamma} \right )^{1/2}
\end{equation}

\subsubsection{ Calculation of $n_a$ } 

The number of particles in the left well A is given by
\begin{equation}
n_a = \int_{-\infty}^{+\infty} dv \; \int_{-\infty}^{+\infty} dx \; P(x,v)
\end{equation}

Since $P(x,v)$ is given by
\begin{eqnarray*}
P(x,v) = \xi(x,v) \; \exp \left [ -\frac{\frac{1}{2} v^2 + V(x)}{k_BT} 
\right ]
\end{eqnarray*}

\noindent
and $\xi(x,v)$ is obtained from Eq.(26). Thus
\begin{eqnarray*}
P(x,v) = F_2 \; \left [ \left ( \frac{\pi\gamma k_BT}{2\lambda} \right )^{1/2}
\; + \; \int_0^u e^{-\frac{\lambda u^2}{2\gamma k_BT} } \; du \right ] \;
e^{ -\frac{\frac{1}{2} v^2 + V(x)}{k_BT} }
\end{eqnarray*}

We have the following condition :

\noindent
$\bullet$ As $x \rightarrow -\infty$, i.e., the left well ; 
$u \rightarrow 
\infty$ [since $u=v+a(x-x_b)$ and $a$ is negative]. Therefore
\begin{eqnarray*}
\int_0^\infty e^{-\frac{\lambda u^2}{2\gamma k_BT} }\; du = 
\sqrt{ \frac{2\pi k_BT\gamma }{2\lambda} }
\end{eqnarray*}
\begin{eqnarray*}
P(x,v) = F_2 \; \left [ \left ( \frac{2\pi \gamma k_BT}{\lambda} \right )^{1/2}
\right ] \; e^{-\frac{\frac{1}{2}v^2 + V(x)}{k_BT} } \; \; .
\end{eqnarray*}

Since
\begin{eqnarray*}
V(x) = V(x_a) + \frac{1}{2} \omega_a^2 (x-x_a)^2
\end{eqnarray*}
\begin{eqnarray}
n_a & = & F_2 \; \left ( \frac{2\pi \gamma k_BT}{\lambda} \right )^{1/2} \;
e^{-\frac{V(x_a)}{k_BT} } \; \int_{-\infty}^{+\infty} e^{-\frac{v^2}{2k_BT} } 
\; dv \int_{-\infty}^{+\infty} e^{- \frac{\omega_a^2 (x-x_a)^2}{2k_BT} } \; dx
\nonumber \\
& = & F_2 \; \left ( \frac{2\pi \gamma k_BT}{\lambda} \right )^{1/2} \;
e^{-\frac{V(x_a)}{k_BT} } \; (2\pi k_BT)^{1/2} \;  \left (
\frac{2\pi k_BT}{\omega_a^2} \right )^{1/2} \nonumber \\
& = & F_2 \; \frac{ (2\pi k_BT)^{3/2} }{\omega_a} \; 
\left ( \frac{\gamma}{\lambda} \right )^{1/2} \; e^{-\frac{V_a}{k_BT} } \; \; .
\end{eqnarray}

We are now in a position to calculate the Kramers' rate
\begin{eqnarray*}
k = j/n_a \; \; .
\end{eqnarray*}

From Eq.(34) and Eq.(36) we get
\begin{equation}
k = \frac{ F_2 \; (k_BT)^{3/2} \; (2\pi)^{1/2} \; \left ( \frac{\gamma}
{\lambda+\gamma} \right )^{1/2} \; e^{-\frac{V(x_b)}{k_BT} } }
{ F_2 \; \frac{ (2\pi k_BT)^{3/2} }{\omega_a} \; \left (
\frac{\gamma}{\lambda} \right)^{1/2} \; e^{-\frac{V(x_a)}{k_BT} } }
\end{equation}
\begin{equation}
k = \frac{\omega_a}{2\pi} \; \left ( \frac{\lambda}{\lambda+\gamma} \right
)^{1/2} e^{- \frac{V(x_b) - V(x_a)}{k_BT} }
\end{equation}

\noindent
where $V(x_b)-V(x_a) = E$, the energy of activation. The pre-exponential
factor in Eq.(35) can be simplified further as follows,

Since $\gamma+\lambda =a_-$ (negative root of $a$ is $a_-$) [see Eq.(17)]
\begin{eqnarray*}
\lambda & = & -(\gamma + a_-) \nonumber \\
& = & - \left [ \gamma + \left \{ -\frac{\gamma}{2} - 
\sqrt{ \left ( \frac{\gamma}{2} \right )^2 + \omega_b^2 } \right \} \; 
\right ] \nonumber \\
& = & -\frac{\gamma}{2} + \sqrt{ \left ( \frac{\gamma}{2} \right )^2
+ \omega_b^2 } \nonumber \\
& = & a_+ \; \; ({\rm positive \; root \; of} \; a)
\end{eqnarray*}

\begin{eqnarray*}
\frac{\lambda}{\lambda+\gamma} = - \left ( \frac{a_+}{a_-} \right ) =
-\frac{a_+ a_+}{a_- a_+} = \frac{ 
\left ( -\frac{\gamma}{2} +
\sqrt{ \left ( \frac{\gamma}{2} \right )^2 + \omega_b^2 } \right )^2 }
{\omega_b^2}
\end{eqnarray*}

We thus get the final expression for Kramers' rate for arbitrary $\gamma$
\begin{equation}
k = \frac{\omega_a}{2\pi\omega_b} \; \left [ -\frac{\gamma}{2} +
\sqrt{ \left ( \frac{\gamma}{2} \right )^2 + \omega_b^2 } \right ] \;
e^{-E/k_BT} \; \; .
\end{equation}

We now consider the two limiting cases in the above equation :

\noindent
$\bullet$ when $\gamma \rightarrow 0$ (i.e., for small viscosity
coefficient)
\begin{equation}
k_{\gamma \rightarrow 0} = \frac{\omega_a}{2\pi} \; e^{-E/k_BT}
\end{equation}
\noindent
which is the transition state result (independent of $\gamma$).

\noindent
$\bullet\bullet$ when $\gamma \rightarrow$ large, i.e., $\gamma \gg \omega_b$
(large viscosity limit)
\begin{eqnarray*}
k_{\gamma \rightarrow {\rm large} } & = & \frac{\omega_a}{2 \pi \omega_b} \;
\left [ \frac{\gamma}{2} \; \left \{ 1+ \frac{1}{2} \; \frac{4\omega_b^2}
{\gamma^2} \right \} - \frac{\gamma}{2} \; \right ] \; e^{-E/k_BT}
\nonumber \\
& = & \frac{\omega_a}{2\pi \omega_b} \; \frac{\omega_b^2}{\gamma} \;
e^{-E/k_BT}
\end{eqnarray*}
\noindent
or
\begin{equation}
k_{\gamma \rightarrow {\rm large} } = \frac{\omega_a \omega_b}{2\pi \gamma} \;
e^{-E/k_BT}
\end{equation}

The rate of reaction is inversely proportional to the viscosity. This 
observation has been corroborated by a number of experimental investigations 
in the recent past.

The general result (39) which is valid in the intermediate to the strong 
damping limit provides a theoretical basis for Arrhenius expression for
reaction rate $k = A e^{-E/k_BT}$ proposed many years ago.

\subsection{A simple connection to Transition State Theory}

We start from Kramers' equation which describes the motion of a particle
in a force field governed by the potential $V(x)$ simultaneously
subjected to Brownian motion :
\begin{eqnarray*}
\frac{\partial}{\partial t} P(x,v,t) = \left [ -\frac{\partial}{\partial x}
v + \frac{\partial}{\partial v} \left \{ V'(x) +\gamma v \right \} +
\gamma k_B T \frac{\partial^2}{\partial v^2} \right ] P(x,v,t)
\end{eqnarray*}

We have shown that the equilibrium distribution $P_{eq}(x,v)$ corresponding
to above equation is given by
\begin{eqnarray*}
P_{eq} (x,v) = Z \; e^{ - \frac{ \frac{1}{2} v^2 + V(x)}{k_BT} }
\end{eqnarray*}

One assumes a that the particles initially residing in the left well are
equilibrated and also that the above distribution is valid for all $x$
( around the bottom of the left well ) except at the barrier top $x=B$. We
therefore put
\begin{eqnarray*}
P_{eq}(x,v) = 0 \; \; \; \; {\rm for} \; \; x>B
\end{eqnarray*}

The normalization constant $Z$ is thus determined by the condition
\begin{eqnarray*}
\int^{+\infty}_{-\infty} dv \int^B_{-\infty} P_{eq}(x,v) \; dx & = & 1 \\
{\rm or}, \; \; Z \int^{+\infty}_{-\infty}
e^{ -\frac{ \frac{1}{2} v^2}{k_BT} } \; dv
\int^B_{-\infty} P_{eq}(x,v) \; e^{ -\frac{V(x)}{k_BT} } \; dx & = & 1 \\
Z \; \sqrt{2\pi k_BT} \; \int^B_{-\infty} e^{ -\frac{V(x)}{k_BT} } \; dx
& = & 1
\end{eqnarray*}

\noindent
Expanding $V(x)$ around the left bottom $x_a$
\begin{eqnarray*}
V(x) = V(x_a) + \frac{1}{2} V''(x_a)(x-x_a)^2
\end{eqnarray*}

\noindent
we have [ write $V''(x_a) = \omega_a^2$ ]
\begin{eqnarray*}
Z \; \sqrt{2\pi k_B T} \; e^{-\frac{V(x_a)}{k_BT} } \; \int^b_{-\infty}
e^{ -\frac{\omega_a^2(x-x_a)^2 }{2k_B T} } \; dx  =  1
\end{eqnarray*}

\begin{eqnarray*}
{\rm or} \; \; Z \; \sqrt{2\pi k_B T} \; \frac{ \sqrt{2\pi k_B T} }{\omega_a} \;
e^{ -\frac{V(x_a)}{k_B T} } = 1
\end{eqnarray*}

\begin{eqnarray*}
{\rm or} \; \; Z = \frac{ \omega_a}{2\pi k_B T } \; 
e^{ \frac{V(x_a)}{k_B T} }
\; \; \; , \; \; \; [ B \rightarrow +\infty ]
\end{eqnarray*}

Therefore the escape rate $k$ over the barrier is obtained by computing the
outward flow over the top of the barrier.
\begin{eqnarray*}
k = \int_0^\infty v\; P(x_b,v) \; dv
\end{eqnarray*}

\noindent
Thus we have
\begin{eqnarray*}
P(x_b,v) = Z \; e^{ -\frac{ \frac{1}{2}v^2 + V(x_b) }{ k_B T} }
\end{eqnarray*}

\noindent
and $k$ as
\begin{eqnarray*}
k & = & Z \; e^{ -\frac{ V(x_b) }{k_B T} } \; \int_0^\infty v
\; e^{ -\frac{v^2}{2 k_B T} } \; dv \\
& = & Z \; e^{ -\frac{ V(x_b) }{k_B T} } \; ( k_B T )
\end{eqnarray*}

\noindent
Putting the value of $Z$ in the above equation we get
\begin{eqnarray*}
k & = & \frac{\omega_a}{2\pi k_B T} \;
e^{ -\frac{ [ V(x_b)-V(x_a) ] }{k_B T} } \; k_B T \\
k & = & \frac{\omega_a}{2\pi} \; e^{-E_0/k_B T}
\end{eqnarray*}

\noindent
where $E_0 = V(x_b)-V(x_a)$ is the activation energy.

This is the transition state result we derived earlier employing Kramers'
method $\gamma = 0$ ( Note that this is not a dynamical theory like that of
Kramers. So $\gamma$ does not appear in the theory. For $\gamma \rightarrow 0$,
one has to consider the problem of energy diffusion. We state the main result;
the rate constant $k$ becomes proportional to $\gamma$ in this limit). 
This result implies that
whenever the Brownian particle is at the top with positive velocity it will
escape as if there is an absorbing wall at the barrier top. A rough
interpretation of this transition state result is that the particle oscillates
in an effective potential $\frac{1}{2}\omega_a^2(x-x_a)^2$ provided
by the left well and therefore hit the wall $\omega_a/2\pi$ times per second
and each time has a probability $e^{-E_0/k_B T}$ to cross over it.

\section{Overdamped motion : Smoluchowski equation and diffusion over a
barrier}

\setcounter{equation}{0}

\subsection{Smoluchowski equation}

We wish to derive an equation of diffusion for an ensemble of particles with 
probability distribution function $f(x,t)$, where the particles in addition 
to Brownian motion execute a deterministic motion in a force field. The 
potential is given by $V(x)$. The evolution of distribution from time $t$ to 
another time $t+\tau$ is given by the following equation
\begin{equation}
f(x, t+\tau) = \int_{-\infty}^{+\infty} f(x-\Delta, t) \; \phi(\Delta) \; 
d\Delta \; \; .
\end{equation}

In presence of Brownian motion, and under the influence of potential $V(x)$
the equation of motion for the particle of unit mass ($m=1$) is given by
\begin{equation}
\ddot{x} + \gamma \dot{x} + V'(x) = F(t)
\end{equation}

\noindent
where $F(t)$ is the random force term.

Under overdamped condition $\ddot{x} \ll \gamma \dot{x}$ we write
\begin{eqnarray}
\gamma \dot{x} & = & -V'(x) + F(t) \\
{\rm or,} \; \; \dot{x} & = & -\frac{V'(x)}{\gamma} + \frac{F(t)}{\gamma}
\; \; .
\end{eqnarray}

The deterministic increment  in $x$ in time $\tau$ corresponding to first 
term on the R.H.S. of Eq.(4) can be calculated as
\begin{equation}
\frac{\partial x}{\partial t} \; \tau = -\frac{V'(x)}{\gamma} \; \tau \; \; .
\end{equation}

Expanding the functions $f$ on both sides of Eq.(1) as usual we have
\begin{eqnarray}
& & f(x,t) + \frac{\partial f}{\partial t} \; \tau + \frac{\partial f}{\partial x}
\left ( \frac{\partial x}{\partial t} \right ) \; \tau \nonumber\\
& = & \int_{-\infty}^{+\infty} \left [ \; f(x,t) -
\frac{\partial f}{\partial x} \; \Delta + \frac{1}{2} \Delta^2 \;
\frac{\partial^2 f}{\partial x^2} \; \right ] \; \phi(\Delta) \; d\Delta
\nonumber \\
& = & f(x,t) \int_{-\infty}^{+\infty} \phi(\Delta) \; d\Delta -
\frac{\partial f}{\partial x} \int_{-\infty}^{+\infty} \Delta \; 
\phi(\Delta) \; d\Delta + \frac{\partial^2 f}{\partial x^2}
\int_{-\infty}^{+\infty} \frac{1}{2} \Delta^2 \; \phi(\Delta) \; d\Delta \; \; .
\end{eqnarray}

\noindent
We now note
\begin{equation}
\left. \begin{array}{ccc} 
\int_{-\infty}^{+\infty} \phi(\Delta) \; d\Delta  & = & 1 \\
\int_{-\infty}^{+\infty} \Delta \; \phi(\Delta) \; d\Delta & = & 0 \\
\frac{1}{2\tau} \; \int_{-\infty}^{+\infty} \Delta^2 \;
\phi(\Delta) \; d\Delta & = & D \end{array} \right \} \; \; .
\end{equation}

Putting Eq.(5) in Eq.(6) and making use of Eq.(7) we get
\begin{eqnarray*}
\frac{\partial f}{\partial t} = - \frac{\partial f}{\partial x} \;
\left (- \frac{V'(x)}{\gamma} \right ) + D\; \frac{\partial^2 f}{\partial x^2}
\end{eqnarray*}

\noindent
or
\begin{equation}
\frac{\partial f}{\partial t} =  \frac{\partial f}{\partial x} \;
\left ( \frac{V'(x)}{\gamma} \right ) + D\; \frac{\partial^2 f}{\partial x^2}
\end{equation}

\noindent
where $D$ is the diffusion coefficient ( position ) and is given by 
$k_B T/\gamma$ as derived earlier ( Einstein's theory ). This equation is 
known as Smoluchowski equation.

\subsection{Diffusion of particles over the barrier}

We consider a particle moving in a potential field $V(x)$ of the type shown
in the Fig.(..). More generally, we may consider an ensemble of particles 
moving in the potential field $V(x)$ without any mutual interference. We
suppose that the particles are initially caught in the potential hole at 
$x_{min}$. The general problem we wish to solve is the rate at which particles
will escape over the potential barrier as a consequence of Brownian motion.

The problem is very complex. However, considerable simplification can be made 
if we assume that the height of the potential barrier is large compared to
the energy of the thermal motions, i.e., $E_0 \gg k_BT$. Under this 
circumstance, the problem can be treated in which the conditions are
{\it quasi-stationary}.

More specifically we may suppose that to a high degree of accuracy an 
equilibrium distribution exists in the neighborhood of $x_{min}$. But this
distribution is not valid for all values of $x$. We assume that beyond
$x_{max}$ the density of particles is very small compared to the equilibrium 
value. And in consequence of this there will be slow diffusion of particles
across $x_{max}$ tending to restore the equilibrium throughout. If the barrier 
were sufficiently high this diffusion will take place as though the stationarity
prevailed. This condition is termed as a quasi-stationary condition.

We thus consider the Smoluchowski equation,
\begin{equation}
\frac{\partial f(x,t)}{\partial t} = \frac{\partial f}{\partial x}
\left ( \frac{V'(x)}{\gamma} \right ) + \frac{k_BT}{\gamma} \;
\frac{\partial^2 f(x,t)}{\partial x^2} \; \; .
\end{equation}

\noindent
Recasting the above equation in the form of a continuity equation we identify
$j$ as the current
\begin{eqnarray}
\frac{\partial f}{\partial t} & = & -\frac{\partial }{\partial x} \left [
- \frac{V'(x) f(x,t)}{\gamma} - \frac{k_BT}{\gamma} \;
\frac{\partial f(x,t)}{\partial x} \right ] \nonumber\\
& = & -\frac{\partial }{\partial x} j(x,t) \; \; .
\end{eqnarray}

\noindent
In the stationary state $j=$constant, i.e., $\frac{\partial f}{\partial t}=0$,
where
\begin{equation}
-j = \frac{ V'(x) f(x,t)}{\gamma} + \frac{k_BT}{\gamma}\; 
\frac{\partial f}{\partial x} \; \; .
\end{equation}

\noindent
Rearranging the above equation as
\begin{equation}
\frac{\partial f(x)}{\partial x} + \frac{V'(x) f(x)}{k_BT} = 
-\frac{j\gamma}{k_BT} \; \; ,
\end{equation}

\noindent
and integrating between $x_{min}$ to $A$ with the integrating factor
\begin{eqnarray*}
e^{+\int \frac{V'(x)}{k_BT} \; dx } \;
\left [ \; = e^{+ \frac{V(x)}{k_BT} } \; \right ]
\end{eqnarray*}

\noindent
the equation (12) in the following form
\begin{equation}
\frac{d}{dx}\left [ \; f(x) \; e^{+ \frac{V(x)}{k_BT} } \; \right ] =
-\frac{j\gamma}{k_BT} \; e^{+ \frac{V(x)}{k_BT} } \; \; ,
\end{equation}

\noindent
we obtain
\begin{equation}
\left [ \; f(x) \; e^{+ \frac{V(x)}{k_BT} } \; \right ]^A_{x_{min}} =
-\frac{j\gamma}{k_BT} \; \int^A_{x_{min}} e^{+ \frac{V(x)}{k_BT} }\; dx \; \; .
\end{equation}

\noindent
The constant current  or flux across $x_{max}$ is
\begin{equation}
j = -\frac{k_BT}{\gamma} \; \frac{
\left [ \; f(x) \; e^{+ \frac{V(x)}{k_BT} } \; \right ]^A_{x_{min}} }{
\int^A_{x_{min}} e^{+ \frac{V(x)}{k_BT} } \; dx } \; \; .
\end{equation}

\noindent
Since $f(x)$ at $A$ is zero, i.e., $f(A)=0$, we have
\begin{equation}
j = \frac{k_BT}{\gamma} \; \frac{
f(x_{min}) \; e^{+ \frac{V(x_{min})}{k_BT} }  }
{ \int^A_{x_{min}} e^{+ \frac{V(x)}{k_BT} } \; dx } \; \; .
\end{equation}

Around $x_{min}$, the current is almost zero. This defines an equilibrium 
condition in the neighborhood of $x_{min}$. Thus with $j=0$ the Smoluchowski 
equation yields ( see Eq.(11) )
\begin{equation}
\frac{k_BT}{\gamma} \; \frac{\partial f}{\partial x} = -\frac{V'(x)}{\gamma}
\; f(x)
\end{equation}
\noindent
or
\begin{eqnarray*}
\frac{\partial f}{\partial x} = -\frac{V'(x)}{k_BT}
\; f(x) \; \; .
\end{eqnarray*}

\noindent
Integrating between $x_{min}$ to $x$ ( a point in the left well )
\begin{eqnarray*}
\ln \frac{f(x)}{f(x_{min})} = -\int^x_{x_{min}} \frac{V'(x)}{k_BT} \; dx
\end{eqnarray*}
\noindent
or
\begin{eqnarray}
f(x) & = & f(x_{min}) \; e^{-\int^x_{x_{min}} \frac{V'(x)}{k_BT} \; dx}
\nonumber\\
& = & f(x_{min}) \; e^{\frac{-V (x)+ V (x_{min}) }{k_BT} }
\end{eqnarray}

\noindent
The equilibrium population in the left well is given by
\begin{equation}
n_a = \int_{x_1}^{x_2} f(x) \; dx =
f( x_{min} ) \int_{x_1}^{x_2} e^{\frac{-V(x) + V(x_{min}) }{k_BT} } \; dx
\end{equation}

\noindent
where $x_1$ and $x_2$ are two points around $x_{min}$.

The rate of escape $k$ is thus given by
\begin{equation}
k = j/n_a \; \; .
\end{equation}

\noindent
Thus from (16) and (19) we obtain

\begin{eqnarray*}
k & = & 
\left ( \frac{k_BT}{\gamma} \right ) \frac{ 
f(x_{min}) \; e^{+\frac{V (x_{min} )}{k_BT} } }{ \int^A_{x_{min}}
e^{+\frac{V(x)}{k_BT} } \; dx} \; \frac{1}{ f( x_{min} ) \int_{x_1}^{x_2}
e^{ \frac{-V(x) + V(x_{min}) }{k_BT} } \;  dx} \nonumber\\
& = & \left ( \frac{k_BT}{\gamma} \right ) \; \frac{
e^{+\frac{V (x_{min} )}{k_BT} } }{ e^{+\frac{V (x_{min} )}{k_BT} } } \;
\frac{1}{
\int^A_{x_{min}} e^{+\frac{V (x)}{k_BT} } \; dx \;
\int_{x_1}^{x_2} e^{-\frac{V (x)}{k_BT} } \; dx } \; \; .
\end{eqnarray*}

\noindent
Therefore
\begin{equation}
k =  \left ( \frac{k_BT}{\gamma} \right ) \; 
\frac{1}{
\int^A_{x_{min}} e^{+\frac{V (x)}{k_BT} } \; dx \;
\int_{x_1}^{x_2} e^{-\frac{V (x)}{k_BT} } \; dx } \; \; .
\end{equation}

We now make use of the following linearization of $V(x)$ around $x_{min}$ and
$x_{max}$. For the integral
\begin{eqnarray*}
\int^A_{x_{min}} e^{+\frac{V (x)}{k_BT} } \; dx
\end{eqnarray*}
\noindent
we use
\begin{equation}
V(x) = E_0 -\frac{1}{2} \; \omega_b^2 \; (x-x_{max})^2
\end{equation}

\noindent
and let $x_{min} \longrightarrow -\infty$ and $A \longrightarrow +\infty$.
Thus
\begin{eqnarray}
\int_{-\infty}^{+\infty} 
e^{+\frac{ [E_0 -\frac{1}{2}\omega_b^2 (x-x_{max})^2 ] }{k_BT} } \; dx
& = & e^{+\frac{ E_0 }{k_BT} } \int_{-\infty}^{+\infty}
e^{-\frac{ (x-x_{max})^2 }{2k_BT/\omega_b^2} } \; dx \nonumber\\
& = & e^{+\frac{ E_0 }{k_BT} } \; 
\frac{ \sqrt{2k_BT} \; \sqrt{\pi} }{\omega_b} \; \; .
\end{eqnarray}

\noindent
For the integral
\begin{eqnarray*}
\int_{x_1}^{x_2} e^{-\frac{V (x)}{k_BT} }\; dx
\end{eqnarray*}

\noindent
we use
\begin{equation}
V(x) = \frac{1}{2}\; \omega_0^2 \; (x-x_{min})^2
\end{equation}

\noindent
and let
\begin{eqnarray*}
\begin{array}{ccc}
x_1 & \longrightarrow & -\infty \\
x_2 & \longrightarrow & +\infty \end{array}
\end{eqnarray*}

\noindent
Therefore
\begin{equation}
\int_{-\infty}^{+\infty} 
e^{-\frac{1}{2} \frac{\omega_0^2}{k_BT} (x-x_{min})^2 } \; dx 
= \frac{ \sqrt{2k_BT} \; \sqrt{\pi} }{\omega_0} \; \; .
\end{equation}

\noindent
Putting the values of these integrals (23) and (25) in the expression for $k$
in Eq.(21) we obtain
\begin{eqnarray}
k & = & \frac{k_BT}{\gamma} \; \frac{ e^{-\frac{E_0}{k_BT} } }{
\left ( \frac{ \sqrt{2k_BT} }{\omega_b} \right ) \; \left (
\frac{\sqrt{2k_BT} }{\omega_0} \right ) \; \pi } \nonumber\\
& = & \frac{ \omega_0 \; \omega_b }{2\; \pi \; \gamma} \;
e^{ -E_0/k_BT } \; \; .
\end{eqnarray}

\noindent
i) The rate of activation has a typical Arhenius form $Ae^{-E_0/k_BT}$.

\noindent
ii) The rate is thus inversely proportional to the friction coefficient of
the medium.

\section{The master equation }

\setcounter{equation}{0}

\subsection{Master equation}

In statistical mechanics we deal with probability distribution functions. The
master equation is a typical probability balance equation.

Let us recall the good old theory of Brownian motion. Suppose a series of 
observations of the same Brownian particle gives a sequence of positions
$x_1$, $x_2$, $x_3$, $\ldots$.

\noindent
Each displacement $x_{k+1}-x_k$ is an element of chance, i.e., is independent
of earlier positions $x_{k-1}$, $x_{k-2}$, etc. This means probability 
distribution function does not depend on the previous history. Thus the position 
$x_{k+1}$ depends only on $x_k$. We call this stochastic process a
{\it Markov Process}. Since many collision have already occurred during the 
displacement $x_{k+1}-x_k$, this displacement is much larger than the
mean free path.

We now recall the basic equation for evolution of probability ( Einstein )
distribution function $f(x,t)$.
\begin{equation}
f(x,t+\tau) = \int f(x+\Delta,t) \; \phi(\Delta) \; d\Delta
\end{equation}

This equation relates the probability distribution function of a Brownian 
particle at $x$ and time $t+\tau$, to that for the particle at a previous
position $x+\Delta$ at an earlier time $t$. $\phi(\Delta)$ is the
probability of a jump of magnitude $\Delta$ ( i.e., $x_{k+1}- x_k$ ) in the 
picture from $x+\Delta$ to $x$.

We now introduce the following notation for convenience. We denote
\begin{equation}
y = x + \Delta
\end{equation}

\noindent
and therefore, $dy = d\Delta$.

\noindent
Also write
\begin{equation}
\phi(\Delta) = \phi(x\rightarrow y) \; \; ,
\end{equation}

\noindent
where the arrow refers to the direction of jump from $x\rightarrow y$. Eqn.(1)
can then be rewritten as [ denote $f(x,t)$ by $P(x,t)$ ]
\begin{equation}
P(x, t+\tau) = \int \phi(y\rightarrow x) \; P(y,t) \; dy \; \; .
\end{equation}

\noindent
Expanding the left hand side as before we write
\begin{equation}
P(x,t) + \tau\frac{\partial P(x,t)}{\partial t} = \int \phi(y\rightarrow x)
\; P(y,t) \; dy \; \; .
\end{equation}

\noindent
Rearranging
\begin{equation}
\tau\frac{\partial P(x,t)}{\partial t} = \int \phi(y\rightarrow x)
\; P(y,t) \; dy - P(x,t) \; \; .
\end{equation}

\noindent
Now note that
\begin{eqnarray}
\int \phi(x\rightarrow y) \; dy & = & \int \phi(-\Delta) \; d\Delta \nonumber\\
& = & 1
\end{eqnarray}
\noindent
(Normalization of probability).

Therefore we write
\begin{equation}
P(x,t) = \int \phi(x\rightarrow y) \; P(x,t) \; dy
\end{equation}

\noindent
Putting (8) in (6) we get
\begin{equation}
\tau\frac{\partial P(x,t)}{\partial t} = 
\int \phi(y\rightarrow x) \; P(y,t) \; dy -
\int \phi(x\rightarrow y) \; P(x,t) \; dy
\end{equation}

\noindent
Dividing both sides by $\tau$ and rewriting
\begin{equation}
W(y \rightarrow x) = \frac{ \phi(y\rightarrow x) }{\tau}
\end{equation}

\noindent
we get
\begin{equation}
\frac{ d P(x,t)}{d t} = 
\int W(y\rightarrow x) \; P(y,t) \; dy -
\int W(x\rightarrow y) \; P(x,t) \; dy
\end{equation}

\noindent
$W(y\rightarrow x)$ is the probability of a jump from  $y\rightarrow x$ per
unit time or the transition probability per unit time.

The above equation is called the master equation. One can immediately write
a discrete version of this equation as [ replace $y$ by $n$, $x$ by $m$ and
integral by summation ]
\begin{equation}
\frac{ d P_m(t) }{ d t } = \sum_n W_{nm} \; P_n(t) -
\sum_n W_{mn} \; P_m(t)
\end{equation}

In this form the meaning of master equation in very clear The first term is 
the gain of state $m$ due to transitions from the other states $n$ and the
second term is the loss due to transitions from $m$ to all other states $n$.
Note that $W_{nm} \geq 0$ when $n \neq m$. The master equation is thus a loss-gain 
equation for probabilities of separate states.

The master equation is a doorway for studying the approach to equilibrium. The
condition for equilibrium is defined as $\frac{d p_m}{dt} =0$, i.e.,
\begin{eqnarray*}
\sum_n W_{nm} \; P_n^{eq} = \sum_n W_{mn} \; P_m^{eq}
\end{eqnarray*}

\noindent
$P_i^{eq} (i=m \; {\rm or} \; n)$ must be identified with the equilibrium
distribution function known from equilibrium statistical mechanics.

The above condition states the fact that {\it in equilibrium} the sum of all
transitions per unit time into any state $m$ must be balanced by the sum of
all transitions from $m$ to all other states $n$. We now state another stronger
condition that for each pair $n,m$ separately the transitions must balance.
\begin{eqnarray*}
W_{nm} \; P_n^{eq} = W_{mn} \; P_m^{eq}
\end{eqnarray*}

This is the principle of detailed balance and is true for all closed, isolated
systems.

\subsection{Applications}

\noindent
{\bf One step process} : If we consider the jumps only between the nearest
neighboring sites then the process is called an one-step process.

The coefficient $r_n$ is the probability/time that being at $n$, a jump 
to $n-1$  has occurred and $g_n$ is probability/time for a jump to $n+1$ from 
$n$. For this
the master equation (12) reduces to
\begin{equation}
\dot{p}_n = g_{n-1}\; p_{n-1} + r_{n+1} \; p_{n+1} - ( r_n p_n + g_n p_n )
\end{equation}

\subsubsection{Example 1 : unidirectional random walk}

Consider a one step process with constant transition probability.
\begin{equation}
r_n = 0 \; \; , \; \; g_n = q
\end{equation}

\noindent
The master equation is
\begin{equation}
\dot{p}_n = q ( p_{n-1} - p_{n} )
\end{equation}

\noindent
It is a random walk over integers $n=0,1,2,\ldots$ with steps only to the 
right at random times. We start with a trial solution
\begin{equation}
p_n = \alpha_n(t) \; e^{-qt} \; \; .
\end{equation}

\noindent
where $\alpha_n(t)$ is to be determined.

\noindent
Hence putting (16) in (15) we obtain
\begin{eqnarray*}
\begin{array}{lll}
\dot{\alpha}_n & = & q \; \alpha_{n-1} \\
\dot{\alpha}_{n-1} & = & q \; \alpha_{n-2} \\
\vdots  & = & \vdots \\
\dot{\alpha}_2 & = & q \; \alpha_1 \\
\dot{\alpha}_1 & = & q \; \alpha_0 \end{array}
\end{eqnarray*}

\noindent
Assume $\alpha_0(t=0)=1$ i.e., $p_0(t=0) =1$ as the initial condition imposed
on (16). Therefore
\begin{eqnarray*}
\begin{array}{lll}
\alpha_1 & = & q \; t \\
\dot{\alpha}_2 & = & q \; \alpha_1 = q^2 \; t \end{array}
\end{eqnarray*}

\noindent
which leads to
\begin{eqnarray*}
\begin{array}{lll}
\alpha_2 & = &  q^2 \; t^2 / 2 \\
\vdots & = & \; \vdots \\
\alpha_n & = & q^n \; t^n / n! \end{array}
\end{eqnarray*}

Therefore the solution is
\begin{equation}
p_n(t) = \frac{ q^n \; t^n }{n!} \; e^{-q \; t}
\end{equation}

which is a Poisson distribution. Note that it is a non-stationary distribution. 
In the next example we consider a wellknown stationary distribution.

\subsubsection{Example 2 : quantized harmonic oscillator interacting with 
a radiation field}

Consider $n=0,1,2,\ldots$ states of a harmonic oscillator. The oscillator
is interacting with a radiation field. The interaction between the oscillator
and the radiation field is causing the transition between the states of the 
oscillator.

\noindent
Energy of a state $n$ is
\begin{equation}
\left ( n + \frac{1}{2} \right ) \; h \; \nu
\end{equation}

\noindent
Since the dipole moment matrix element between $n-1$ and $n$ is proportional  
to $n$, probability for a jump from  $n-1 \rightarrow n$
\begin{equation}
g_{n-1} = \beta \; n
\end{equation}

\noindent
Probability of a jump from $n \rightarrow n-1$
\begin{equation}
r_n = \alpha \; n
\end{equation}

\noindent
$\alpha$, $\beta$ are dependent on the frequency of light. We start from
\begin{eqnarray}
\dot{p}_n & = & g_{n-1} \; p_{n-1} + r_{n+1} \; p_{n+1} - (g_n + r_n)\; 
p_n \nonumber\\
\dot{p}_n & = & \beta \; n \; p_{n-1} + \alpha(n+1) \; p_{n+1} - 
[\beta(n+1) + \alpha \; n ] \; p_n
\end{eqnarray}

\noindent
For a stationary distribution $\dot{p}_n=0$. Therefore we have
\begin{eqnarray*}
0 = \beta \; n \; p_{n-1} + \alpha(n+1) \; p_{n+1} - 
[\beta(n+1) + \alpha \; n ] \; p_n
\end{eqnarray*}

\begin{equation}
\beta \; (n+1) \; p_n - \alpha(n+1) \; p_{n+1} =
\beta \; n \; p_{n-1} - \alpha \; n  \; p_n  = {\rm const} = 0 \; \; .
\end{equation}

\noindent
Hence
\begin{eqnarray*}
\beta \; p_{n-1} = \alpha \; p_n
\end{eqnarray*}

\noindent
Therefore
\begin{eqnarray*}
p_n = \frac{\beta}{\alpha} \; p_{n-1}
\end{eqnarray*}

\noindent
which we rewrite as
\begin{equation}
p_n = \left ( \frac{\beta}{\alpha} \right )^n \; p_0 \; \; \; , \; \; \;
p_0 = {\rm constant}
\end{equation}

\noindent
This distribution is an equilibrium distribution. We know from equilibrium 
statistical mechanics that
\begin{equation}
p_n = {\rm const} \times e^{ -nh\nu/kT }
\end{equation}

\noindent
Therefore equating (23) and (24) we obtain
\begin{eqnarray*}
\left ( \frac{\beta}{\alpha} \right )^n & = &
\left ( e^{ -h\nu/kT } \right )^n \nonumber  \\
{\rm or } \; \; \frac{\beta}{\alpha} & = &  e^{ -h\nu/kT }
\end{eqnarray*}

\noindent
Since $g_n$ is proportional to the  radiation density $\rho$ present
\begin{eqnarray*}
\beta = C \; \rho
\end{eqnarray*}

\noindent
For $r_n$ which is given by $\alpha$ there are spontaneous ($A$) and
stimulated processes ($B\rho$),
\begin{eqnarray*}
\alpha = B\rho + A
\end{eqnarray*}

\noindent
Therefore we have
\begin{eqnarray*}
\frac{A+B\rho}{C\rho} = e^{ -h\nu/kT }
\end{eqnarray*}

\noindent
Rearranging we write
\begin{equation}
\rho = \frac{ A }{ C e^{ -h\nu/kT } - B }
\end{equation}

\noindent
which has the form of a Planck's distribution. $A$, $B$ and $C$ can be
determined by comparing with Rayleigh-Jeans law and the Wein's law in long 
wavelength and short wavelength limits.

%%%%%%%%%%%%%%%%%%%%%%%%%%%%%%%%%%%%%%%%%%%%%%%%%%%%%%%%%%%%%%%%%%%%%%%%%%%%%

%%%%%%%%%%%%%%%%%%%%%%%%%%%%%%%%%%%%%%%%%%%%%%%%%%%%%%%%%%%%%%%%%%%%%%%%%%%%%

\begin{appendix}

\section{Evaluation of Eq.(17)}

\begin{eqnarray*}
f(x,t) & = & \frac{n}{2\pi} \int_{-\infty}^{+\infty} e^{ikx} \; e^{-k^2 Dt} 
\; dk \nonumber \\
& = & \frac{n}{2\pi} \; e^{-x^2/4Dt} \int_{-\infty}^{+\infty} e^{ikx} \;
e^{(k\sqrt{Dt})^2} \; e^{x^2/4Dt} \; dk \nonumber\\
& = & \frac{n}{2\pi} \; e^{-x^2/4Dt} \int_{-\infty}^{+\infty}
\exp \left [ - \left \{ (k\sqrt{Dt})^2 + (-ix/2\sqrt{Dt})^2 + (-ikx) \right \}
\right ] \; dk \nonumber\\
& = & \frac{n}{2\pi} \; e^{-x^2/4Dt} \int_{-\infty}^{+\infty}
\exp \left [ - (k\sqrt{Dt} - (ix/2\sqrt{Dt} ) )^2 \right ] \; dk \nonumber\\
& = & \frac{n}{2\pi} \; e^{-x^2/4Dt} \int_{-\infty}^{+\infty}
\frac{1}{\sqrt{Dt}} \; e^{-y^2} \; dy \; \; ; \; \; {\rm put} \; 
y=k\sqrt{Dt} - (ix/2\sqrt{Dt}), \; dy = \sqrt{Dt} \; dk \nonumber\\
& = & \frac{n}{2\pi\sqrt{Dt}} \; e^{-x^2/4Dt} \; \sqrt{\pi} \nonumber\\
f(x,t) & = & \frac{n}{\sqrt{4\pi Dt}} \; e^{-x^2/4Dt}  \; \; .
\end{eqnarray*}

This result shows that a Fourier transform of a Gaussian function is a 
Gaussian.

\end{appendix}

%%%%%%%%%%%%%%%%%%%%%%%%%%%%%%%%REFERENCES%%%%%%%%%%%%%%%%%%%%%%%%%%%%%%%%%%

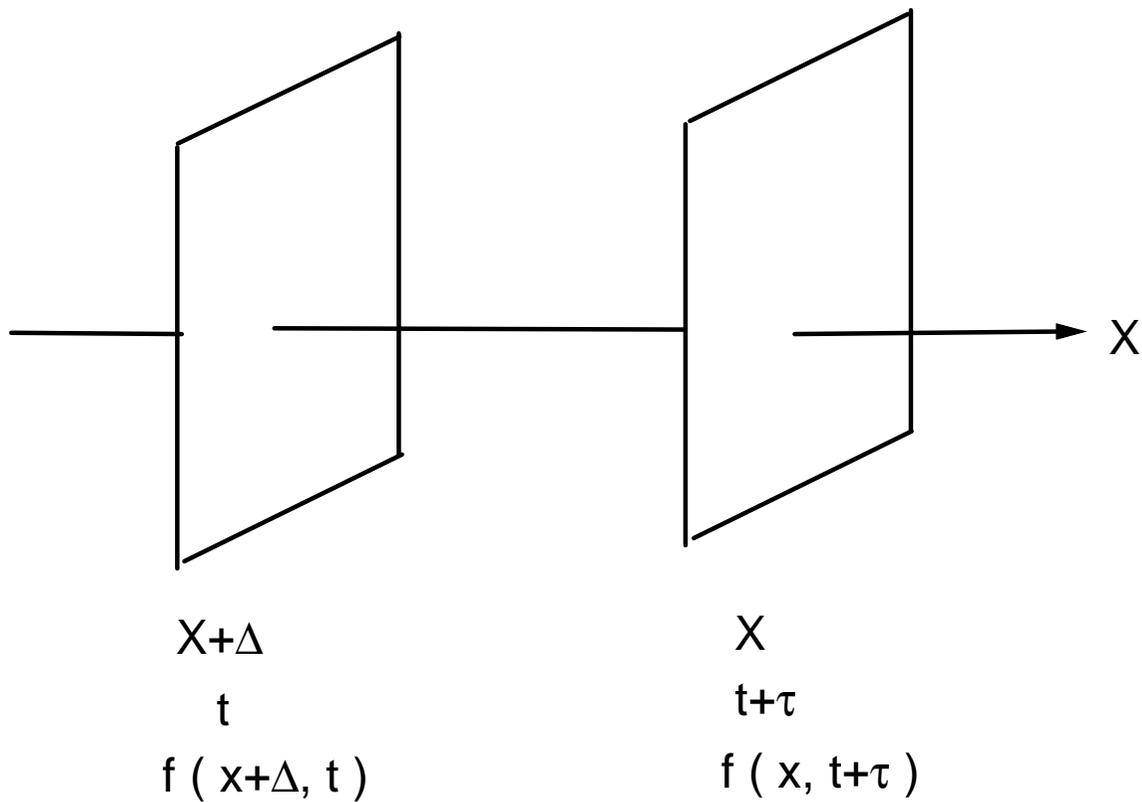
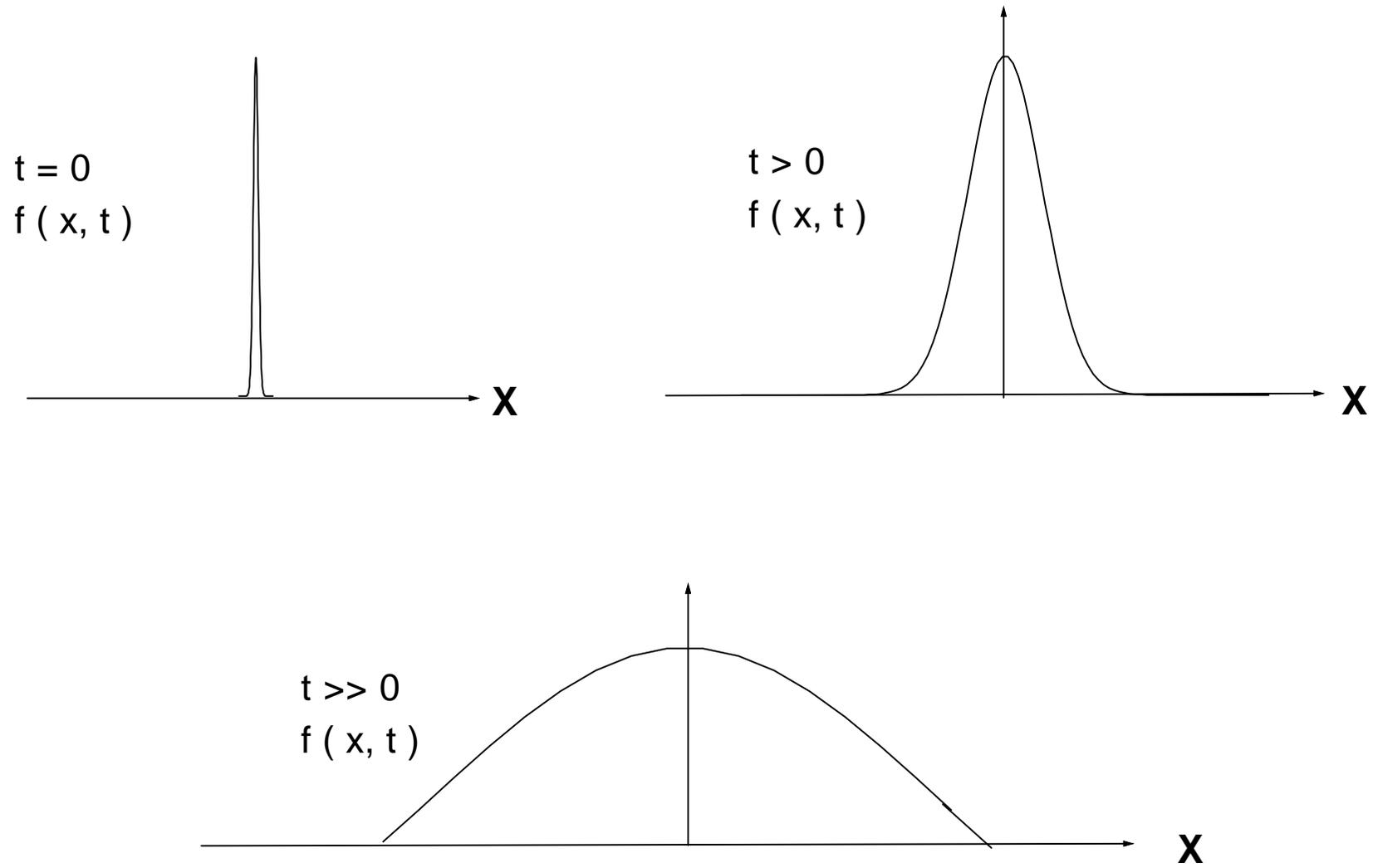
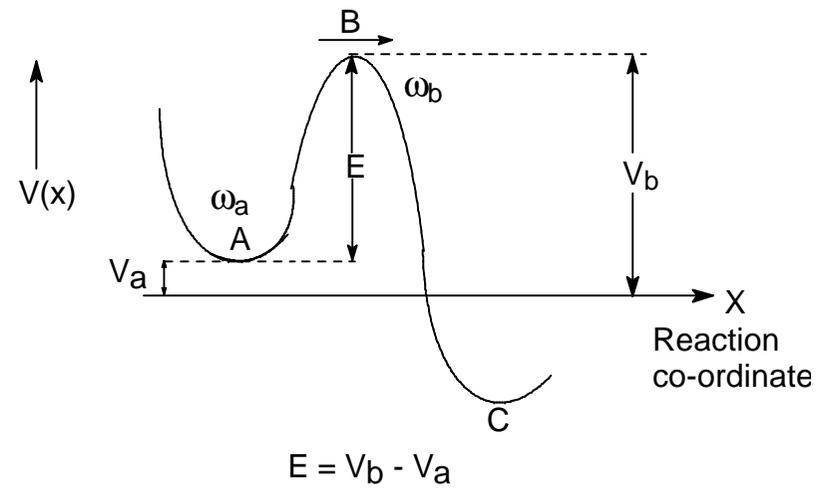
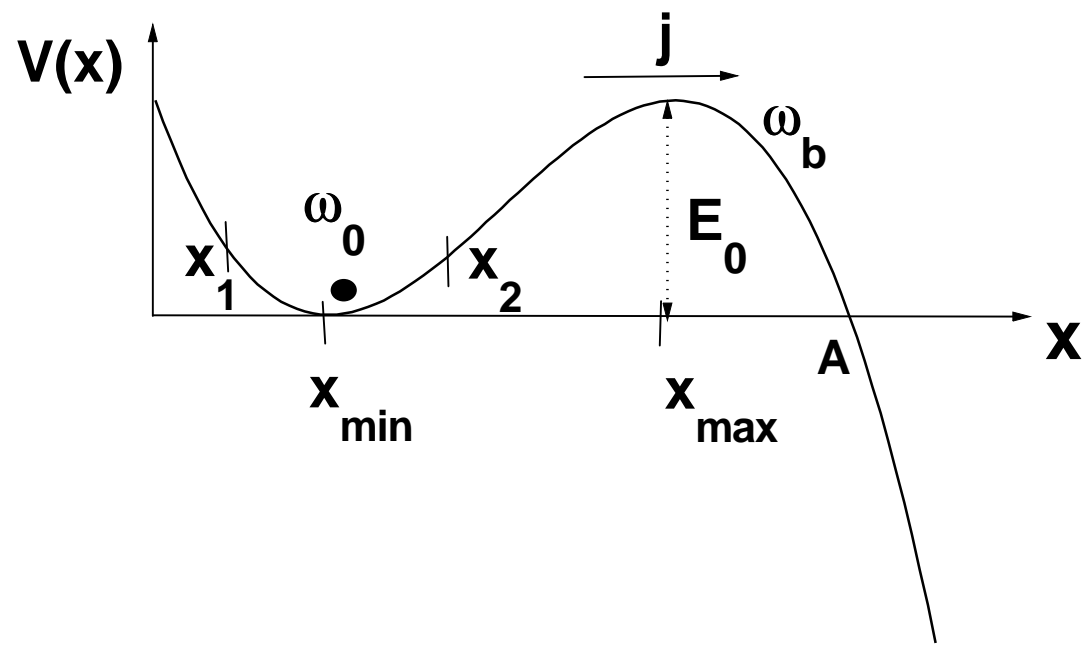
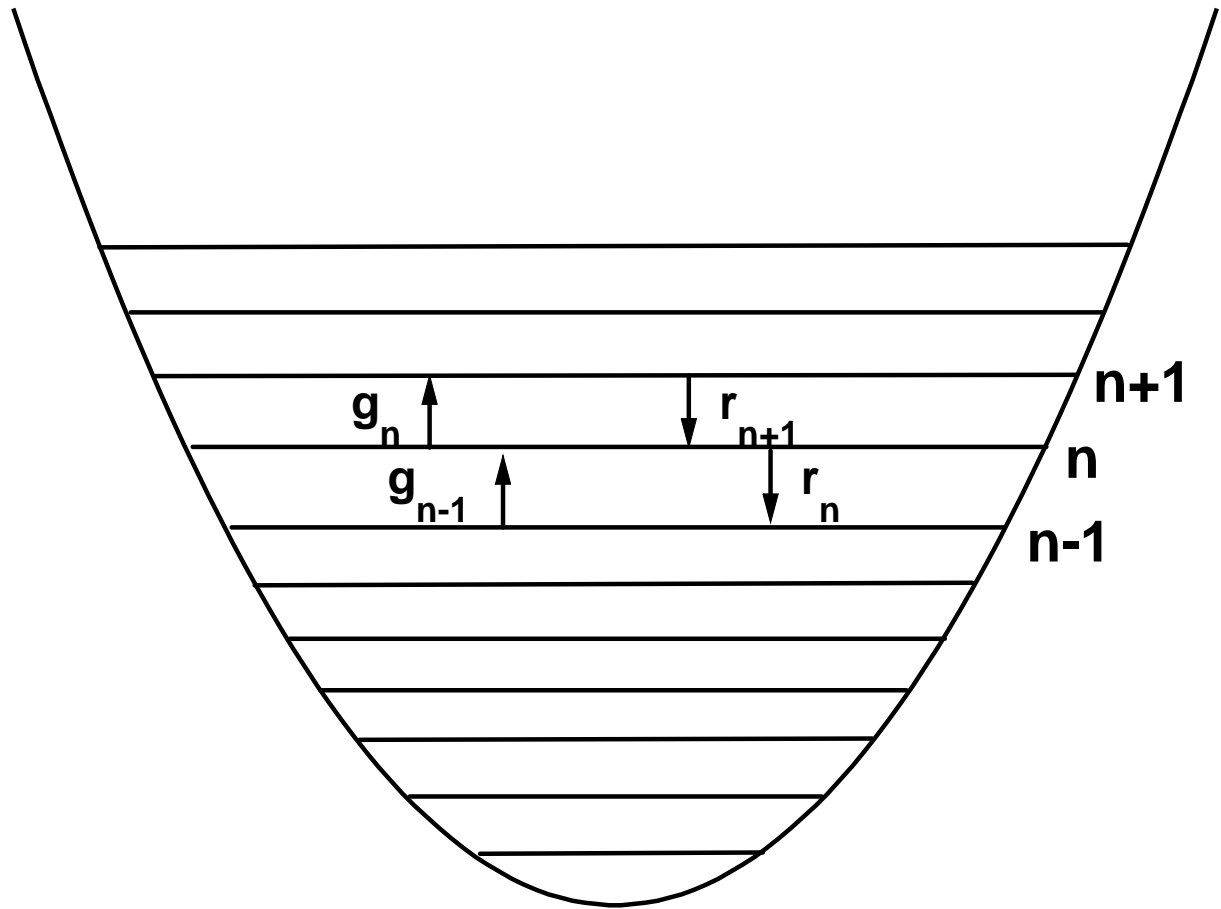
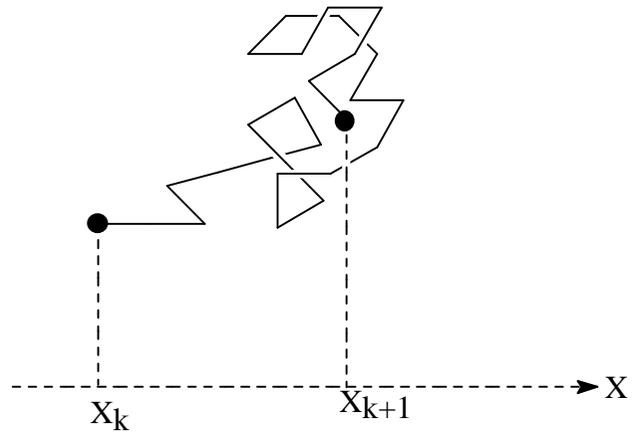
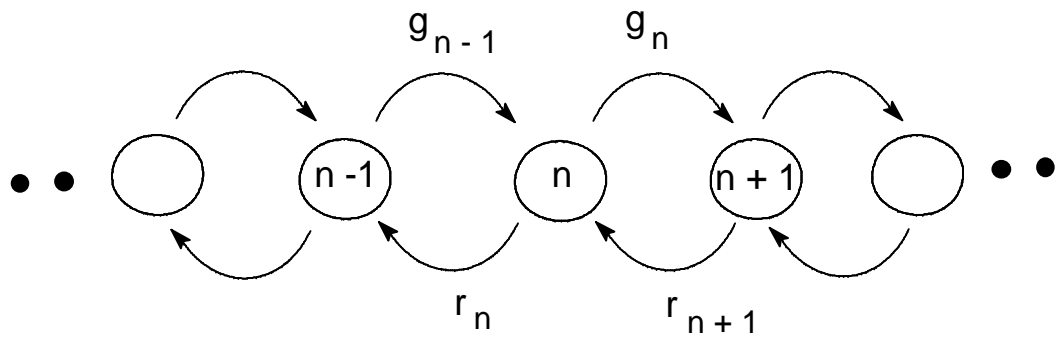
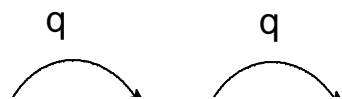
\end{document}